\documentclass[12pt]{amsart}
\usepackage{amsmath, amssymb, amsthm, amsfonts, amsthm, bm, eucal, graphicx}

\theoremstyle{plain}

\newtheorem{thm}{Theorem}[section]
\newtheorem{cor}[thm]{Corollary}
\newtheorem{lem}[thm]{Lemma}
\newtheorem{prop}[thm]{Proposition}

\theoremstyle{definition}
\newtheorem{defi}[thm]{Definition}
\newtheorem{defis}[thm]{Definitions}
\newtheorem{conj}[thm]{Conjecture}
\newtheorem{conv}[thm]{Convention}
\newtheorem{nota}[thm]{Notation}
\newtheorem{rem}[thm]{Remark}
\newtheorem{rems}[thm]{Remarks}
\newtheorem{exa}[thm]{Example}
\newtheorem{exas}[thm]{Examples}
\newtheorem{sit}[thm]{}

\newcommand{\brem}{\begin{rem}}
\newcommand{\brems}{\begin{rems}}
\newcommand{\erem}{\end{rem}}
\newcommand{\erems}{\end{rems}}
\newcommand{\bexa}{\begin{exa}}
\newcommand{\bexas}{\begin{exas}}
\newcommand{\eexa}{\end{exa}}
\newcommand{\eexas}{\end{exas}}
\newcommand{\bdefi}{\begin{defi}}
\newcommand{\edefi}{\end{defi}}
\newcommand{\bdefis}{\begin{defis}}
\newcommand{\edefis}{\end{defis}}
\newcommand{\bcor}{\begin{cor}}
\newcommand{\ecor}{\end{cor}}
\newcommand{\blem}{\begin{lem}}
\newcommand{\elem}{\end{lem}}
\newcommand{\bconv}{\begin{conv}}
\newcommand{\econv}{\end{conv}}
\newcommand{\bconj}{\begin{conj}}
\newcommand{\econj}{\end{conj}}
\newcommand{\bprop}{\begin{prop}}
\newcommand{\eprop}{\end{prop}}
\newcommand{\bthm}{\begin{thm}}
\newcommand{\ethm}{\end{thm}}
\newcommand{\bnota}{\begin{nota}}
\newcommand{\enota}{\end{nota}}
\newcommand{\bsit}{\begin{sit}}
\newcommand{\esit}{\end{sit}}
\newcommand{\be}{\begin{eqnarray}}
\newcommand{\ee}{\end{eqnarray}}
\newcommand{\bproof}{\begin{proof}}
\newcommand{\eproof}{\end{proof}}

\def\ba{\begin{array}}
\def\ea{\end{array}}

\def\cF{{\mathcal F}}

\def\cE{{\mathcal E}}

\newcommand{\card}{\operatorname{card}}

\newcommand{\spec}{\operatorname{spec}}
\newcommand{\Mat}{\operatorname{Mat}}
\newcommand{\Res}{\operatorname{Res}}

\newcommand{\id}{\operatorname{id}}

\newcommand{\Aut}{{\operatorname{Aut}}}

\newcommand{\Irr}{{\operatorname{Irr}}}
\newcommand{\ford}{{\operatorname{ford}}}
\newcommand{\sord}{{\operatorname{sord}}}
\newcommand{\adj}{{\operatorname{adj}}}
\newcommand{\Tr}{{\operatorname{Tr}}}

\newcommand{\veg}{{\operatorname{\cF}(\G, k)}}

\newcommand{\Fix}{{\operatorname{Fix}}}
\newcommand{\GF}{{\operatorname{GF}}}
\newcommand{\vect}{{\operatorname{vect}}}
\newcommand{\lcm}{{\operatorname{lcm}}}

\newcommand{\GL}{{\bf {GL}}}

\renewcommand{\div}{{\operatorname{div}}}

\newcommand{\A}{{\mathbb A}}

\newcommand{\R}{{\mathbb R}}
\newcommand{\C}{{\mathbb C}}

\newcommand{\Z}{{\mathbb Z}}
\newcommand{\N}{{\mathbb N}}
\newcommand{\NO}{{\mathbb N}_{\rm odd}}
\newcommand{\NE}{{\mathbb N}_{\rm even}}
\newcommand{\T}{{\mathbb T}}
\newcommand{\F}{{\mathbb F}}
\newcommand{\G}{{\Gamma}}
\newcommand{\VG}{{\operatorname{vert(\G)}}}
\newcommand{\ver}{{\operatorname{vert}}}
\newcommand{\ord}{{\operatorname{ord}}}

\newcommand{\EG}{{\operatorname{edg (\G)}}}
\newcommand{\edg}{{\operatorname{edg}}}
\newcommand{\roots}{{\operatorname{roots}}}
\newcommand{\Harm}{{\operatorname{Harm}}}
\newcommand{\End}{{\operatorname{End}}}

\newcommand{\D}{{\Delta}}

\begin{document}

\title[Periodic binary harmonic functions]
{Periodic binary harmonic functions on
lattices}

\author{Mikhail Zaidenberg}
\address{Universit\'e
Grenoble I, Institut Fourier, UMR 5582 CNRS-UJF, BP 74, 38402
St.\ Martin d'H\`eres c\'edex, France}
\email{zaidenbe@ujf-grenoble.fr}

\thanks{
{\bf Acknowledgements:} This article was written  partially during the
author's visit to the Max-Planck-Institute of Mathematics at Bonn. 
The author thank this
institution for generous support.}

\thanks{
\mbox{\hspace{11pt}}{\it 1991 Mathematics Subject
Classification}: 11B39, 11T06, 11T99, 31C05, 37B15, 43A99.\\
\mbox{\hspace{11pt}}{\it Key words}: Chebyshev-Dickson
polynomial, Fibonacci polynomial, cellular automaton, graph
laplacian, spectrum of a graph, binary function, discrete harmonic
function, pluri-periodic function, finite abelian group, lattice,
elliptic curve, characteristic two.}



\begin{abstract}
A function  on a (generally infinite) graph $\G$ with values 
in a field $K$ of
characteristic $2$ will be called {\it  harmonic} if its value at
every vertex of $\G$ is the sum of its values over all adjacent
vertices. We consider binary pluri-periodic harmonic functions 
$f: \Z^s\to\F_2=\GF(2)$
on integer lattices, and address the problem of describing 
the set of possible multi-periods $\bar n=(n_1,\ldots,n_s)\in\N^s$ of such
functions. Actually this problem arises in the
theory of cellular automata \cite{MOW, Su1, Su4, GKW}. 
It occurs to be
equivalent to determining, for a certain
affine algebraic hypersurface $V_s$ in $\A_{\bar\F_2}^s$, the torsion
multi-orders of the points on $V_s$ in the multiplicative group
$(\bar\F_2^\times)^s$. In particular $V_2$ is an elliptic cubic curve.
In this special case we provide a more thorough treatment. A major
part of the paper is devoted to a survey of the subject. 
\end{abstract}

\maketitle

{\footnotesize \hskip 2.0in {\it In mathematics, our role is more
that of servant than master.}

\hskip 2.0in Charles Hermite, from reminiscences  by Jacques
Hadamard.}

\smallskip

{\footnotesize \tableofcontents}

\section*{Introduction}
We use the following notation. We let $\G= (\VG, \EG)$ be a (not
necessarily finite) graph without loops and multiple
edges, $K$ be a field of characteristic $2$,
$k=\F_2=\GF(2)$ the {\it binary field}; moreover,
we write $\bar k = \bar F_{2}$ for its algebraic closure. 

For $K$-valued functions $f: \VG\to K$, we consider two kinds of laplacians:
$\Delta^+_\G$ is averaging  over balls of radius $1$, 
respectively $\Delta^-_\G$ is averaging  over 
spheres of radius $1$ (cf. e.g., \cite{Ca, Ba}) so that
$$\Delta^+_\G=\id +\Delta^-_\G, \qquad\mbox{where}
\qquad \Delta^-_\G (f)(v)=
\sum_{[v,v']\in\,\EG} f(v')\,,\quad v\in\VG\,.$$ Actually
$\Delta^\pm$ is the $\sigma^\pm$-{\it cellular
automaton}
studied  e.g. 
by Martin, Odlyzko and Wolfram \cite{MOW}, Sutner \cite{Su1}-\cite{Su4}, 
Goldwasser,
Klostermeyer and Ware \cite{GKW}, Sarkar, Barua and Ramakrishnan 
\cite{BR, SB}, Hunzikel, Machiavello and Park \cite{HMP} e.a. 

A function $f$ on $\VG$ is called {\it harmonic} ({\it antiharmonic}) if 
$ \Delta^+_\G (f)= 0$
($ \Delta^-_\G (f)= 0$, respectively). 
Thus for $f$ harmonic, $f(v)$ is equal to the sum
of values of $f$ over the neighbors of $v$, whereas for $f$
antiharmonic this sum is always zero. We let 
$$\operatorname{Harm}_K^+(\G)=\ker\, (\Delta^+_\G)\qquad\mbox{ and}
\qquad \operatorname{Harm}_K^-(\G)=\ker\, ( \Delta^-_\G)=\ker\, (
\id+\Delta^+_\G)$$ be the corresponding subspaces of the vector space
$\cF(\G, K)$ of all $K$-valued functions on $\VG$. We simply write
$\operatorname{Harm}^\pm(\G)=\operatorname{Harm}_k^\pm(\G)$
when dealing with binary functions.

The support $N^\pm=N^\pm_\G (f)=\operatorname{supp} (f)$ of a nonzero binary
harmonic (antiharmonic) function $f$ will be called
a {\it nucleus} ({\it antinucleus}, respectively) of $\G$.
Note that the constant function $1$ is harmonic on an
odd graph and antiharmonic on an even one. More generally, the
(anti)nuclei can be characterized by the following two properties:
\begin{enumerate} \item[$\bullet$] Every nucleus $N^+$ of
$\G$ is an odd subgraph, that is each vertex of $N^+$ is of odd
degree within $N^+$. Whereas every antinucleus $N^-_\G$ is an
even subgraph. \item[$\bullet$] Every vertex $v\in \VG\setminus
\operatorname{vert} (N^\pm_\G)$ has an even number of neighbors
in $ N^\pm_\G$. \end{enumerate}

\noindent Nuclei are {\it even-parity} subgraphs of $\G$,
see e.g. Amin, Slater and Zhang \cite{ASZ}, Klostermeyer \cite{K} 
and literature therein on parity
domination in graphs. The set of all (anti)nuclei in
$\G$ is closed under symmetric difference.

\bdefi \label{binuc} We say that $\G$ is {\it harmonic}
({\it antiharmonic}, respectively) if there exists a nonzero binary
harmonic (antiharmonic, respectively) function on $\G$. A {\it
biharmonic} graph is a graph that is simultaneously harmonic and
antiharmonic.\edefi

A locally finite graph $\G$ will be called {\it even} ({\it odd}, respectively)
if the degree of every vertex $v$ of $\G$ is even (odd, respectively).
Every odd graph is harmonic and every even one is
antiharmonic. Moreover, any  (anti)harmonic graph can be obtained
from an odd (respectively, an even) one by adjoining a certain
number of new vertices, each one being joined with the old
ones by an even number of edges. 

For instance, an infinite plane hexagonal lattice is odd and therefore
harmonic, whereas an infinite plane triangular lattice 
is even and so antiharmonic.
Actually the latter one is biharmonic, the nuclei being the maximal
inscribed hexagonal lattices. Similarly every integer
lattice $\Z^s$ is biharmonic when regarded as a graph with edges 
parallel to the
coordinate axes. 
In all these examples, none of the
(anti)nuclei is finite. 
However it is easy to find infinite graphs with
finite nuclei.

We let $\spec^\pm (\G)$ be the spectrum of the laplacian
$\Delta^\pm_\G$ in the algebraic closure $\bar k$. Since the
laplacian $\D^\pm_\G$ is defined over the binary field $k$, we
have
  $$d^\pm:=\dim_k (\Harm^\pm (\G))
  =\dim_{\bar k} (\Harm^\pm_{\bar k}(\G))\,.$$
  Thus by definition, $\G$ is harmonic if and only if $ 0\in
\spec^+ (\G)$, antiharmonic if and only if $ 0\in \spec^- (\G)$,
and biharmonic if and only if $ 0, 1\in \spec^+ (\G)$. The shift
by $ 1$ being an involution on $\bar k$, this justifies our
terminology.

Our aim is to determine the set of all multi-indices $\bar
n=(n_1,\ldots,n_s)\in\N^s$ such that the integer lattice
$\Z^s$ possesses an $\bar n$-periodic nonzero binary harmonic
function. For instance, for $\bar n= (1,\ldots, 1)$ such a
function $f$ on $\Z^s$ must be constant $\equiv  1$, but the
constant function $1$ on $\Z^s$ is not harmonic, although it is
antiharmonic.

Given a Galois covering $\pi:\G'\to\G$ with the Galois group $G$, there is 
an isomorphism
$$\pi^*: \operatorname{Harm}^\pm(\G)\stackrel{\cong}
{\longrightarrow} [\operatorname{Harm}^\pm(\G')]^G\, ,\qquad
f\longmapsto f\circ\pi,$$ where the latter vector 
space consists of all $G$-stable
(anti)harmonic functions on $\G'$. Thus our  problem reduces to
the following one: 

Letting $C_n$
be a circular graph with $n$ vertices, we consider 
a finite abelian group $\Z_{\bar
n}:=\prod_{i=1}^s \Z/n_i\Z$ and 
the associated toric lattice (the Caley graph of $\Z_{\bar
n}$)
$$\T_{\bar n}=C_{n_1}\times\ldots\times C_{n_s},\qquad \bar
n=(n_1,\ldots,n_s)\in\N^s\,.$$  Then $\Z^s$ can be
viewed as the maximal abelian cover of $\T_{\bar n}$ with the
Galois group $G=\sum_{i=1}^s n_i\Z \vec e_i\subseteq \Z^s$.
Hence the space $[\operatorname{Harm}^+ (\Z^s)]^G$ of
pluri-periodic binary harmonic functions on $\Z^s$ with periods
$n_1\vec e_1,\ldots,n_s\vec e_s$ can be identified with
$\operatorname{Harm}^+(\T_{\bar n})$. So we would like to
determine the set of all harmonic toric lattices $\T_{\bar n}$.

In section 1 we deal with (anti)harmonic functions on trees. 
Following Amin, Slater and Zhang \cite{ASZ}, 
Gravier, Mhalla and Tanner \cite{GMT} 
we provide in \ref{fore} 
an algorithm that computs the dimension $d^\pm$ 
of the vector space 
$\operatorname{Harm}^\pm (\G)$. 

In
section 2 we give an account for some spectral properties 
of laplacians on multidimensional grids and toric
lattices. We
also mention some uniqueness theorems for binary harmonic functions on
graphs and an application to the game 'Lights Out`. 
One of the typical results of this section is as
follows (see \ref{prod2}.b). 

\bthm\label{resum}  Given $\bar
n=(n_1,\ldots,n_s)\in\N^s$, there exists a nonzero $\bar n$-periodic binary
harmonic function on $\Z^s$ (that is, 
the toric lattice $\T_{\bar n}$ is harmonic) if and only if 
the affine variety in
$\A^s_{\bar k}$ with equations \be\label{ME} \sum_{i=1}^s
(x_i+x_i^{-1})= 1,\qquad x_i^{n_i}= 1,\quad i=1,\ldots,s\,, \ee is
nonempty.
\ethm

Section 3 is devoted to 2-dimensional grids and lattices. According to the above theorem and 
taking into
account the covering trick, to distinguish the harmonic tori it is enough to determine 
all bi-torsions $(\ord\,x,\ord\,y)\in\NO^2$ \footnote{We denote by $\ord\,\xi\in\NO$ the
multiplicative order of an element $\xi\in\bar k^\times$.} 
of points $(x,y)$ on the elliptic cubic curve $E^*$
in $\A^2_{\bar k}$ with equation
$$x+1/x+y+1/y=1\,.$$ As suggested by Zagier, we consider an equivalence 
relation on the set $\NO$ of
all odd natural numbers defined by the connected components of the \
{\it partnership graph}
${\mathcal P}^{(1)}$ with the set of vertices $\NO$ and the set of edges 
$$\{[\ord\,x,\ord\,y]\,:\,(x,y)\in
E^*\}\,.$$
We indicate some simple properties of this graph. 
In particular, all its connected components are
finite (Theorem \ref{MT}). In Appendix 1
the reader will find an overview of the first 13 connected components of 
${\mathcal P}^{(1)}$ calculated by
Zagier with PARI. 

Finally in
Appendix 2 we provide a survey on binary Chebyshev-Dickson and
Fibonacci polynomials, as these are closely related to our subject. 

We are grateful to Don Zagier for his clarifying suggestions in
section 3 and the calculations in Appendix 1. Our thanks also
to Roland Bacher, Silvain Gravier, Lenny Makar-Limanov and Pieter Moree for
highly useful discussions and pointing out references, 
to Andrzej Schinzel for disproving a conjectural
inequality for the Euler function in \ref{gn}.2,  to Gottfried Barthel and 
Roland Bacher
for helpful editorial remarks, and to Andrey Inshakov for his
assistance with MAPLE. Gottfried Barthel also helped with some computations 
for the Euler
function. 

The aim of the present survey being rather pedagogical, 
we have to apologize that the list of references
is by no means complete, 
neither our survey follows the chronological order of
events.

\section{Harmonic forests}
The following proposition is well known, see e.g., \cite{An},
\cite[4.1-4.3]{Su2}, cf. also \cite{DG}, \cite{OZ}. For the sake
of completeness, we give a simple argument that applies in the
characteristic 2 case.

For a vertex $v\in\ver (\G)$ (respectively, for an edge $[u,v]\in
\EG$) we let $\G-v$ (respectively, $\G-[u,v]$) be the graph
obtained from $\G$ by deleting $v$ and all incident edges
(respectively, the edge $[u,v]$ but not the vertices $u$ and
$v$). We let $\adj (\G)$ be the
adjacency matrix of a finite graph $\G$. Notice that $\adj (\G)$ coincides
with the matrix of the laplacian $\Delta^-_\G$. We let
$\chi_\G^{\ } (x)$ be the characteristic polynomial of $\adj
(\G)$, and denote  by $\left(\frac{e}{i}\right)_\G$ the number of
$i$-matchings in $\G$ i.e., of all possible choices of $i$
non-incident edges among the $e$ edges of $\G$.

\bprop\label{grdet} For any finite graph $\G$ with $n$ vertices
and $e$ edges, the following hold:

\begin{enumerate}\item[(a)] $\chi_\G^{\ }(x)=\sum_{i=0}^{[n/2]}
\left(\frac{e}{i}\right)_\G x^{n-2i}$. In particular $n$ and $\chi_\G^{\
}$ are of the same parity.
\item[(b)] $\forall v\in\ver (\G)$,
$$\chi_\G^{\ }(x)= x\!\cdot\!\chi_{\G-v}^{\ }(x)+\sum_{[v,v']\in\edg (\G)}
\chi_{\G-\{v,v'\}}^{\ }(x)\,.$$
\item[(c)] $\forall [u,v]\in\edg (\G)$,
$$\chi_\G^{\ }(x)= \chi_{\G-[u,v]}^{\ }(x)+\chi_{\G-\{u,v\}}^{\ }(x)\,.$$
\item[(d)] $\forall [u,v]\in\edg (\G)$ with $\deg u=1$,
$$\chi_\G(x)= x\!\cdot\!\chi_{\G-v}^{\ }(x)+\chi_{\G-\{u,v\}}^{\ }(x)\quad
\mbox{and so,}\quad
\chi_\G^{\ }(0)=\chi_{\G-\{u,v\}}^{\ }(0)\,.$$
\item[(e)] Given $u,v,w\in\ver (\G)$ such that $\deg
u=\deg v=1$ and $[u,w],\,[v,w]\in\edg (\G)$ (that is $u,v$
are extremal vertices of $\G$ joint with $w$) one has
$$\chi_\G^{\ }(x)= x^2\!\cdot\!\chi_{\G-\{u,v\}}^{\ }(x)
\quad\mbox{and so,}\quad
\chi_\G^{\ }(1)= \chi_{\G-\{u,v\}}^{\ }(1)\,.$$
\item[(f)] Given $u,v,w\in\ver (\G)$ such that $\deg
u=1,\,\deg v=2$ and $[u,v],\,[v,w]\in\edg (\G)$ (that is
$[u,v]$ is an extremal edge of $\G$ joint with $w$)  one has
$$\chi_\G^{\ }(x)= (1+x^2)\!\cdot\!\chi_{\G-\{u,v\}}^{\ }(x)+
x\!\cdot\!\chi_{\G-\{u,v,w\}}^{\ }(x)\quad\mbox{and so,}\quad
\chi_\G^{\ }(1)= \chi_{\G-\{u,v,w\}}^{\ }(1)\,.$$
\end{enumerate} \eprop

\bproof The order $n$ symmetric determinant $\det =\sum_{\sigma\in
S_n} m_{\sigma}$, where as usual $S_n$ stands for the $n$-th 
symmetric group and
$m_{\sigma}=\pm\prod_{i=1}^n
a_{i,\sigma (i)}$, reduces modulo $2$ to $\sum_{\sigma^2={\rm
id}} m_{\sigma}$ by cancelling equal terms $m_{\sigma}$ and
$m_{\sigma^{-1}}$ with $\sigma^2\neq {\rm id}$. This leads to
(a). Now (b) and (c) can be easily deduced from (a). In turn (d),
(e) and (f) can be deduced by virtue of (b). \eproof

\bsit\label{redu} Let  $\G$ be a finite forest that is, a disjoint
union of trees. It can be reduced, in two different ways, to a
rather simple one, by \begin{enumerate} \item[$\bullet$] 
iteratively suppressing
an extremal vertex (leaf) as in (d).
In this way we finally reduce $\G$ to a forest $\G_{\rm
red}^{-}$ with only isolated vertices;
\item[$\bullet$] 
iteratively suppressing a
pair of extremal vertices as in (e) or a
pair of extremal edges as in
(f). Via this procedure, $\G$ will be finally reduced to a forest $\G_{\rm
red}^{+}$ with only isolated vertices
and isolated edges. \end{enumerate}
This gives the following result, see
\cite{ASZ} or, in any positive characteristic, \cite[Theorem 4
and Corollary 6]{GMT}. We recall the notation $d^\pm (\G)=\dim
(\Harm^\pm (\G))$.\esit

\bcor\label{fore} \begin{enumerate}\item[(a)] A forest $\G$ is
harmonic (antiharmonic, respectively) if and only if $\G_{\rm red}^{+}$ 
($\G_{\rm red}^{-}$, respectively) 
contains
an isolated edge (an isolated vertex, respectively).
\item[(b)] Moreover, for any $\G_{\rm red}^{+}$ ($\G_{\rm red}^{-}$,
respectively), the number of isolated
edges (of isolated vertices, respectively) is $d^+ (\G)$ ($d^- (\G)$,
respectively).
\end{enumerate}\ecor

\bproof (a) For a disjoint union $\G=\G'\cup\G''$ of two graphs we have
$\chi_{\G}^{\ }=\chi_{\G'}^{\ }\chi_{\G''}^{\ }$. Since
$\det(\D^-_\G)=\chi_{\G}^{\ }(0)$ and $\det
(\D^+_\G)=\chi_{\G}^{\ }(1)$, we get $\det (\D^\pm_\G)= \det
(\D^\pm_{\G'})\det (\D^\pm_{\G''})$. By virtue of (d)-(f), the
first reduction preserves $\chi_{\G}^{\ }(0)$, and the second one
$\chi_{\G}^{\ }(1)$, so that $\det (\D^\pm_\G)=\det
(\D^\pm_{\G_{\rm red}^\pm})$. Thus $\det (\D^-_\G)=1$ if and only
if $\G_{\rm red}^-$ is empty, and  $\det (\D^+_\G)=1$ if and only
if $\G_{\rm red}^+$ consists of isolated vertices. This proves
(a).

(b) Following our iterative procedure, we can easily see that
every (anti)harmonic function on $\G$ restricts to a (anti)harmonic function
on $\G_{\rm red}^+$ (on $\G_{\rm red}^-$, respectively). Moreover we can
reconstruct the (anti)harmonic functions on $\G$ from their
restrictions to $\G_{\rm red}^\pm$, respectively. Indeed, for any isolated
vertex $v$ of $\G_{\rm red}^-$, the $\delta$-function $\delta_v$ on $\ver
(\G_{\rm red}^-)$, which takes value $1$ at $v$ and $0$ at any
other vertex, is antiharmonic. At every step, $\delta_v$
uniquely extends from a smaller graph to a bigger one 
preserving antiharmonicity. 
This results finally in an antiharmonic
function $\tilde\delta_v$ on $\G$. 

On the other hand, given an
antiharmonic function $h$ on $\G$, it is uniquely determined by
the restriction $h\mid\G_{\rm red}^-$. 
This restriction can be decomposed in
the basis of $\delta$-functions $\left(\delta_v\,:\, v\in\ver
(\G_{\rm red}^-)\right)$ in $\cF (\G_{\rm red}^-, k)$. Hence
$\left(\tilde\delta_v\,:\, v\in\ver (\G_{\rm red}^-)\right)$ form a
basis of $\Harm^- (\G)$.

Similarly, given an isolated edge $[u,v]$ of $\G_{\rm red}^+$,
$\delta_{[u,v]}=\delta_u+\delta_v$ is a harmonic function on
$\G_{\rm red}^+$. At each step it extends uniquely 
to a harmonic function on a bigger graph, and finally
to a function $\tilde \delta_{[u,v]}\in\Harm^+ (\G)$. 
These functions form a basis of $\Harm^+ (\G)$. 
This shows (b).
\eproof

\brem\label{extfor} The analysis of
(anti)harmonicity of unicyclic graphs can be reduced in the same way to that of
cyclic graphs $C_n$ \cite[\S 4]{Su2}. 
As for the latter one, see section 2.1 below. \erem

\section{
Chebyshev-Dickson-Fibonacci polynomials and harmonicity}

\subsection{1-dimensional case} 
We refer the reader to Appendix 2 for a survey 
on the Chebyshev-Dickson polynomials $T_n$
($E_n$) of the first (second) kind and the Fibonacci
polynomials $F_n$.  
We also need the following notation.

\bsit\label{subord} For $n\in\NO$, the order and the suborder of
2 modulo $n$ are, respectively,
$$f(n)=\ord_n\, 2=\min\{j \,:\, 2^j\equiv 1 \!\!\mod  n\}$$ and
$$ f_0(n)=\sord_n\, 2=\min\{j\,:\, 2^j\equiv
\pm 1 \!\!\mod  n\}\,.$$ Thus $f(n)/f_0 (n)\in\{1,2\}$. Moreover,
$$f(n)= 2 f_0(n)\,\,\,\iff \,\,\, \exists j\in\N\,:\, 2^j\equiv
-1\!\! \mod  n \,.$$
Letting $q=2^{f_0 (n)}$ we note that $n\mid (q-1)$ if $f_0
(n)=f(n)$ and $n\mid (q+1)$ otherwise. Anyhow, $n$ divides exactly
one of $q-1$ and $q+1$. Further, $f (2^r-1)=f_0(2^r-1)=r$
$\forall r\ge 3$ (but $f_0(3)=1, \,f(3)=2$) and $f_0
(2^r+1)=r=f(2^r+1)/2$ $\forall r\ge 1$, see Appendix B in
\cite{MOW}. \esit

\bsit\label{charpo} We notice that $\Delta^-_{C_n}=\tau +
\tau^{-1}$, where $\tau\in\End (\A_k^n)$ is the cyclic right shift,
and $\Delta^-_{P_n}=\tau_l + \tau_r$, where $\tau_l$ ($\tau_r$)
$\in\End (\A_k^n)$ is the left (right) shift. Hence the adjacency
matrices of the graphs $C_n$ and $P_n$ are, respectively,
$$\adj (C_n)=\left(\begin{array}{ccccccc}
0 & 1 & 0 & \ldots & 0 & 0 & 1\\
1 & 0 & 1 & \ldots & 0 & 0 & 0\\
0 & 1 & 0 & \ddots & 0 & 0 & 0\\
\vdots & \vdots & \ddots  & \ddots & \ddots  & \vdots & \vdots \\
0 & 0 & 0 & \ddots & 0 & 1 & 0\\
0 & 0 & 0 & \ldots & 1 & 0 & 1\\
1 & 0 & 0 & \ldots & 0 & 1 & 0
 \end{array}\right),\quad
 \adj (P_n)=
 \left(\begin{array}{ccccccc}
0 & 1 & 0 & \ldots & 0 & 0 & 0\\
1 & 0 & 1 & \ldots & 0 & 0 & 0\\
0 & 1 & 0 & \ddots & 0 & 0 & 0\\
 \vdots & \vdots  & \ddots  & \ddots & \ddots & \vdots  & \vdots \\
0 & 0 & 0 & \ddots & 0 & 1 &  0\\
0 & 0 & 0 & \ldots & 1 & 0 &  1\\
0 & 0 & 0 & \ldots & 0 & 1 & 0
 \end{array}\right)\,
$$
with the characteristic polynomials $\chi_{C_n}^{\ }=T_n$ and
$\chi_{P_n}^{\ } =E_n$, respectively. 
For $n$ odd we have
$\spec (\tau)= \mu_n$, where $\mu_n$ stands for the cyclic group of
$n$-th roots of unity in $\bar k$.
According to the spectral mapping theorem, $\spec
(\Delta^-_{C_n})=\{\xi+\xi^{-1}\,:\, \xi\in\mu_n\}$ and $\spec
(\Delta^+_{C_n})=\{1+\xi+\xi^{-1}\,:\, \xi\in\mu_n\}$. Moreover,
for the circular graphs $C_n$ the following results hold, see e.g. 
\cite[4.1, 6.1]{Su3}, \cite[2.1]{SB},
\cite[3.3.8]{HJ}.\esit

\bprop\label{dpo0} 
\begin{enumerate}\item[(a)] $\forall
n\ge 3$, $C_n$ is antiharmonic (that is $\Delta^-_{C_n}$ is
non-invertible). Whereas $C_n$ is harmonic (i.e., $\Delta^+_{C_n}$
is non-invertible) if and only if $n\equiv 0 \!\!\!\mod 3$.
\item[(b)] The minimal polynomial of $\Delta^-_{C_n}$ is
$F_{k}$ if $n=2k$ and $xR_k$, where $R_k=\sqrt{F_n}$, if $n=2k+1$. 
\item[(c)]
The polynomial $xR_k$ having simple roots, for every $n\in\NO$
the matrix $\adj (C_n)$ is similar over $\bar k$ to the diagonal
matrix ${\rm diag} (\zeta^i+\zeta^{-i}\,:\, i=0,\ldots,n)$, where
$\zeta\in\mu_n$ is a primitive $n$-th root of unity.
\item[(d)] Consequently, $\forall n\in\NO$,
$(\Delta^\pm_{C_n})^q=\Delta^\pm_{C_n}$, where $q=2^{f_0(n)}$.
\item[(e)] The kernel of $\Delta^+_{C_{3k}}$ is two-dimensional,
spanned by the
vector $(1,1,0,1,1,0,\ldots)$ and its shift. If $n$ is even then
the kernel of $\Delta^-_{C_{n}}$ is also two-dimensional, spanned
by the vector $(1,0,1,0,1,0,\ldots)$ and its shift. For $n$ odd
this kernel is one-dimensional, spanned by $(1,1,1,1,1,\ldots)$.
Hence $d^+ (C_{3k})=2=d^- (C_{2k})$ and $d^- (C_{2k-1})=1$
$\forall k\ge 1$.
\item[(f)] Respectively,
the nuclei of $C_{3k}$ are the cyclic shifts of $N^+=\{v_i\,:\,
i\not\equiv 0 \!\mod 3\}$, the antinuclei of $C_{2k}$ are the
cyclic shifts of $N^-=\{v_i\,:\, i\not\equiv 0 \!\!\mod 2 \}$,
whereas $N^-=C_n$ is the only antinucleus of $C_n$, $\forall
n=2k-1$.
\end{enumerate}\eprop

Similarly, for the paths $P_n$ we have the following results, 
see e.g. \cite{MOW}, \cite{Su3}, \cite[4.4]{BR},
\cite[3.3-3.4]{SB}.

\bprop\label{pera} 
\begin{enumerate}\item[(a)] $\exists
(\Delta^-_{P_{n-1}})^{-1}\quad\iff\quad n\in\NO$, and $\exists
(\Delta^+_{P_{n-1}})^{-1}\quad\iff\quad n\not\equiv 0 \!\!\mod
3$.
\item[(b)] The
minimal polynomial of $\Delta^-_{P_n}$ is $E_n$.
\item[(c)] $\forall n\in\NO$, $\Delta^-_{P_n}$ 
admits a generalized inverse
$\kappa_{P_n}\in\End (\cF(P_n, k))$ such that
$\Delta^-_{P_n}\kappa_{P_n}\Delta^-_{P_n}=\Delta^-_{P_n}$.
\item[(d)] $\forall n\in \NO$,
$\ord (\Delta^-_{P_{n-1}})=2e_{n}-2$, where 
$e_{n}=\min\{j\in\N\,:\,
(\Delta^-_{C_{n}})^j=\Delta^-_{C_{n}} \}$, $e_n\in\NE$ is such that 
$(e_{n}-1)\mid (q-1)$ for $q=2^{f_0(n)}$.
\item[(e)] $\forall n\in \NO$,
$(\Delta^+_{P_{n-1}})^{2q}=(\Delta^+_{P_{n-1}})^2$. Furthermore,
if $n\in\NO$ and $n\not\equiv 0 \!\!\!\mod 3$ then  $\ord
(\Delta^+_{P_{n-1}})\mid (2q-2)$, where $q=2^{f_0(n)}$.
\item[(f)]  The only nucleus of $P_{3k-1}$ is $N^+=\{v_i\,:\,
i\not\equiv 0 \!\mod 3\}$, and the only antinucleus of $P_{2l-1}$
is $N^-=\{v_i\,:\, i\equiv 1 \!\mod 2 \}$. Hence $d^+
(P_{3k-1})=d^- (P_{2l-1})=1$ $\forall k,l\in N$.
\end{enumerate}\eprop

\brem\label{bib} For every $n\in \NO$, both
$(\Delta^-_{P_{n-1}})^{-1}$  and the generalized inverse
$\kappa_{P_{n}}$ for $\Delta^-_{P_{n}}$ are explicitly found in
\cite{SB}. \erem

\subsection{Spectra of products}
\bsit\label{saba} Letting $E,E'$ be vector spaces over a field
$K$ and $e_1,\ldots,e_m$ ($e'_1,\ldots,e'_n$, respectively) be a
basis of $E$ ($E'$, respectively), we represent every
$X=\sum_{i,j} x_{i,j}e_i\otimes e'_j\in E\otimes E'$ by the
matrix (or {\it pattern}) $X=(x_{i,j})\in\Mat_{m,n}(K)$. Following
\cite[\S 5]{BR}, \cite[\S 4]{SB}, for any two square matrices
$A\in \Mat_{m,m}(K)$ and $B\in\Mat_{n,n}(K)$ we consider the
Sylvester derivation
$$\delta_{A,B}\in\End (E\otimes E'),\quad X\longmapsto AX+X B^t,
\qquad\mbox {with the matrix}\quad C=A\otimes 1+1\otimes B\,.$$
\esit

\noindent The following lemma is well known \cite[VIII.3]{Ga} and
holds for arbitrary fields. We provide a simple argument in the
characteristic $2$ case for $K=\bar k$.

\blem\label{bach} (a) In the notation of \ref{saba} we have
$$\chi_C^{\ } (x)=\Res_y \left(\chi_A^{\ } (x+y), \chi_B^{\ }
(y)\right)\quad\mbox{and}\quad
\spec (C)=\spec (A)+\spec (B)$$ (the Minkowski sum  in $\bar k$ \footnote{We recall 
that for a commutative
semigroup $\Pi$ and for two subsets $\Lambda_1,\Lambda_2\subseteq \Pi$ 
their Minkowski sum is 
$\Lambda_1+\Lambda_2=\{\lambda_1+\lambda_2\,:\,\lambda_i\in\Lambda_i,
\,i=1,2\}$.}).

(b) $C$ is invertible if and only if the characteristic
polynomials $\chi_A^{\ }$ and $\chi_B^{\ }$ are coprime. \elem

\bproof Let $A=S_A+N_A$ be the Jordan decomposition of $A\in\End
(\A_{\bar k}^n)$, with $S_AN_A=N_AS_A$, where 
$S_A, N_A\in\End (\A_{\bar k}^n)$, $S_A$
is semi-simple and $N_A$ is nilpotent. Then
$A^q=S_A^q=S_A$ for certain $q=2^r,\,r>0$.

We fix $q=2^r$ so that $A^q=S_A,\,\,B^q=S_B$ and $C^q=S_C$. Since
$A\otimes 1$ and $1\otimes B$ commute, we have $$S_C=C^q:
X\longmapsto A^qX+X(B^t)^q=S_AX+X S_{B}\,$$ i.e., $S_C=S_A\otimes
1 + 1\otimes S_B$. If the bases $(e_i)$, $(e_j')$ as in
\ref{saba} are diagonalizing for $S_A$, $S_B$, respectively, with
$S_A (e_i)=\lambda_i e_i$ and $S_B (e_j')=\mu_j e_j'$, then
$(e_i\otimes e'_j)$ is a diagonalizing basis for $S_C$ with $S_C
(e_i\otimes e'_j)= (\lambda_i+\mu_j)e_i\otimes e'_j$. For any two
polynomials $p=\prod_{i=1}^m (x+\lambda_i)$ and $q=\prod_{j=1}^m
(x+\mu_j)$ we have \cite{vdW}
$$\Res_y \left (p(x+y), q(y)\right)=\prod_{1\le i\le m,1\le j\le n}
\left(x+\lambda_i+\mu_j\right)\,.$$ Since $\chi_A^{\
}=\chi_{S_A}^{\ }$ etc., the assertions follow easily. \eproof

\brem\label{annih} If $C^q=S_C$ then $(p(C))^q=p(C^q)=p(S_C)$
$\forall p\in \bar k [x]$. It follows that $p_{\rm min}^q (C)=0$,
where $$ p_{\min} (x)=\prod_{\gamma=\lambda+\mu,\,\lambda\in \spec
(A),\,\mu\in\spec (B)} (x+\gamma)\,.
$$\erem

\bsit\label{produ} The Cartesian product $\G=\G_1\times\G_2$ of
two graphs $\G_1,\G_2$ is defined via
$$\operatorname{vert}(\G)=\operatorname{vert}(\G_1)\times
\operatorname{vert}(\G_2),\quad\operatorname{edg}(\G)
=[\operatorname{vert}(\G_1) \times\operatorname{edg}(\G_2)]\cup
[\operatorname{vert}(\G_2)\times\operatorname{edg}(\G_1)]\,.$$ In
particular, the $m\times n$-{\it grid} is the product
$P_{m,n}=P_m\times P_n$,  and the {\it toric $m\times n$-lattice}
is the product $T_{m,n}=C_m\times C_n$.

Fixing an ordering of the $m$ ($n$) vertices of $\G_1$ ($\G_2$),
we may regard any $K$-valued function on $\G_1\times\G_2$ as an
$m\times n$-matrix $X$ with entries in $K$. The laplacian
$\Delta^\pm_\G$ acts on $X$ via
$$\Delta^\pm_{\G}:
X\longmapsto \adj (\G_{1})^\pm\cdot X + X\cdot \adj  (\G_{2})=\adj
(\G_{1})\cdot X + X\cdot \adj  (\G_{2})^\pm\,,
$$ where $A^-=A$ and $A^+=A+1$.
 \esit

 We let $\spec^\pm (\G)=\spec (\D^\pm_\G)\subseteq \bar k$
 and  $\chi_\G^{\ }=\chi_{\adj (\G)}^{\ }$.
 From \ref{bach}  we deduce such a corollary, see e.g. \cite[Lemma 8]{Ba}.

 \bcor\label{bache} \begin{enumerate}\item[(a)]
 The spectrum $\spec^- (\G)$ of the product $\G=\G_1\times\G_2$
 of two graphs is the Minkowski sum of the spectra $ \spec^- ({\G_i})$,
 $i=1,2$. Moreover
 $$\chi_\G^{\ } (x)=\Res_y \left(\chi_{\G_1}^{\ } (x+y),\chi_{\G_2}^{\ }
 (y)\right)\,.$$
 Whereas $$
 \spec^+ (\G)= 1+ \spec^- ({\G_1})+\spec^- ({\G_2})\,.$$
 \item[(b)] Consequently, $\G$ is antiharmonic if and only if the
 characteristic polynomials
 $\chi_{\G_1}^{\ },\,\chi_{\G_2}^{\ }$ 
 are not coprime, and is harmonic
 if and only if the polynomials $\chi_{\G_1}^{\ },\,\chi_{\G_2}^{+}$ 
 are not coprime.
 \end{enumerate}\ecor

\subsection{2-dimensional grids and tori} 
The following is immediate from
\ref{bache}, see \cite{BR}, \cite{Su3}, \cite{HMP}.

\bprop\label{dpll}  \begin{enumerate}
\item[(a)] The grid $P_{m-1,n-1}$ is antiharmonic
(respectively, harmonic) if and only if the Chebyshev-Dickson
polynomials $E_{m-1}$ and $E_{n-1}$ (respectively, $E_{m-1}$ and
$E_{n-1}^+$) are not coprime.
Furthermore $P_{m-1,n-1}$ is antiharmonic if and
only if $\gcd (m, n)\neq 1$.
\item[(b)] $\det(\D^+_{P_{m-1,n-1}})=\Res_x (E_{m-1}, E_{n-1}^+)$.
\item[(c)] $\forall m,n\ge 3$, the toric lattice
$\T_{m,n}$ is antiharmonic and, moreover, is an even graph.
Furthermore $\T_{m,n}$ is harmonic if and only if the polynomials
$T_m$ and $T^+_n$ are not coprime.
\end{enumerate} \eprop

\bcor\label{q-1} \begin{enumerate}\item[(a)] $\forall k,l\in\N$,
the grids $P_{2k-1,3l-1}$ and $P_{3k-1,2l-1}$ are harmonic. The
grid $P_{m-1,n-1}$ different from any one of these is harmonic if
and only if the system \be\label{ssy}
u+u^{-1}+v+v^{-1}=1=u^m=v^n\ee admits a solution $(u,v)\in (\bar
k^\times)^2$.
\item[(b)] We have $$\spec
(\D^-_{\T_{m,n}})=\{u+u^{-1}+v+v^{-1}\,:\,
u\in\mu_m,\,v\in\mu_n\}\,,$$ respectively, $$\spec
(\D^+_{\T_{m,n}})=\{1+u+u^{-1}+v+v^{-1}\,:\,
u\in\mu_m,\,v\in\mu_n\}\,.$$ Thus $\T_{m,n}$ is harmonic if and
only if the system (\ref{ssy}) admits a solution, if and only if either
$mn\equiv 0 \!\!\mod  3$ or $P_{m-1,n-1}$ is harmonic.
\item[(c)] $\forall m,n\equiv
0 \!\!\mod  5$, both the grid $P_{m-1,n-1}$ and the toric lattice
$\T_{m,n}$ are harmonic.
\item[(d)] If $\T_{m,n}$, respectively, $P_{m-1,n-1}$
is harmonic then so is $\T_{km,ln}$, respectively, $P_{km-1,ln-1}$
$\forall k,l\in\N$.
\item[(e)]
$\forall q=2^a, \,\forall q'=2^b$, $\T_{m,n}$ is harmonic if and
only if $\T_{qm,q'n}$ is, and $P_{m-1,n-1}$ is harmonic if and
only if either $P_{qm-1,q'n-1}$ is, or one of the following holds:
$m\equiv 0 \!\!\mod  2 ,\,n\equiv 0  \!\!\mod  3$ or $m\equiv 0
\mod 3,\,n\equiv 0
 \!\!\mod  2 $.
\item[(f)] In particular $\forall a,b\ge 0$,
$\T_{q,q'}$ and $P_{q-1,q'-1}$ are not harmonic.
\end{enumerate}\ecor

\bproof $P_{m-1,n-1}$ is harmonic if and only if
$E_{m-1}(z)=E_{n-1}(z+1)=0$ for some $z\in\bar k$. These
equations are satisfied by $z=0$ (respectively, $z=1$) 
if and only
if $m\equiv 0 \!\!\mod  2 ,\,n\equiv 0  \!\!\mod  3$
(respectively, $m\equiv 0
 \!\!\mod  3,\,n\equiv 0  \!\!\mod  2 $), see \ref{dpo11}.a,e. 
 Suppose further
that $z\neq 0,1$. Letting
$$z=u+u^{-1},\,z+1=v+v^{-1},\quad\mbox{where}
\quad u,v\in\bar k^\times\,,$$
by virtue of (\ref{dpe}) in \ref{dpo} and \ref{dpo11}.a we obtain
$$E_{m-1}(z)=E_{m-1} (u+u^{-1})=0 = E_{n-1}^+ (z)=E_{n-1}
(v+v^{-1})$$ $$\iff\quad T_m(u+u^{-1})=u^m+u^{-m}=0=
T_n(v+v^{-1})=v^n+v^{-n}\,.$$ This shows (a). The same argument
proves (b). The assertions (c), (d) and (e) follow from (b) by
virtue of \ref{dpo11}.e and, in turn, imply (f). \eproof

 In order to find all  harmonic toric
lattices it is enough, by virtue of \ref{q-1}.e, to restrict to
$\T_{m,n}$ with $(m,n)\in\NO^2$. The following facts are established in 
\cite[Theorem 14]{GKW}, see also \cite[5.1]{HMP}.

\bprop\label{dppl}
$\forall q=2^r$, $r\ge 1$, the toric lattices $\T_{q-1,q-1},\quad
\T_{q-1,q+1}$ and $\T_{q+1,q+1}$ are harmonic except for
$\T_{1,1}$ and $\T_{7,7}$. \eprop

\bproof Letting, according to \ref{dpo1}.c,d, 
$$A_q=\roots (T_{q-1})=\{0\}\cup \{z\in
\F_q^\times\,:\, \Tr_{\F_q} (z^{-1})=0\}\,,$$
$$B_q=\roots
(T_{q+1})=\{0\}\cup \{z\in \F_q^\times\,:\, \Tr_{\F_q}
(z^{-1})=1\}
$$ and
$$A_q^+=\roots (T_{q-1}^+)=1+A_q,\qquad B_q^+=\roots
(T_{q+1}^+)=1+B_q\,$$ we have $A_q\cup
B_q=\F_q$ and $A_q\cap B_q=\{0\}$. Indeed by \ref{dpo1}.g, 
$T_{q+1}+T_{q-1}= x^{q+1}$ and
$T_{q+1}T_{q-1}= x^{2q}+x^2=x^2(x^{q-1}+1)^2$. So the zeros of the product
$T_{q+1}T_{q-1}$ fill in $\F_q$, while $0$ is the only common zero of 
$T_{q+1}$ and $T_{q-1}$. 
Hence $\card (A_q)=q/2$ and
$\card (B_q)=q/2 +1$. It follows that $A_q\cap
B_q^+\neq\emptyset$ and $B_q\cap B_q^+\neq\emptyset$. Thus the
polynomials $T_{q-1},\,T_{q+1}^+$, respectively,
$T_{q+1},\,T_{q+1}^+$ are not coprime. In view of \ref{dpll}.c,
the toric lattices $\T_{q-1,q+1}$ and $\T_{q+1,q+1}$ are harmonic
$\forall q=2^r,\,r\ge 1$.

Suppose further that $r\ge 2$ and $\T_{q-1,q-1}$ is not harmonic,
that is $A_q\cap A_q^+=\emptyset$. Then $A_q^+\subseteq
B_q\setminus \{0\}$. Actually $A_q^+= B_q\setminus \{0\}$ as
these sets have the same cardinality. Thus $\roots
(T_{q-1}^+)=\roots (F_{q+1})$. More precisely,
$$F_{q+1}=(x+1)T_{q-1}^+\quad\iff\quad
x^q+F_{q-1}=(x+1)^2 F_{q-1}^+ \quad\iff\quad x^q+1=F_{q-1}^+
+x^2F_{q-1}\,.$$ For every $z\in\F_q\setminus\F_2$ we obtain
$z+1=F_{q-1} (z+1)+z^2F_{q-1} (z)$. Equivalently, by virtue of
\ref{dpo11}.d, \be\label{z+1} (z+1) \left(1+\Tr_{\F_q}
((z+1)^{-1})\right)=z^3\Tr_{\F_q} (z^{-1})\,.\ee From (\ref{z+1})
we deduce the following alternative.

\smallskip

\noindent $\bullet$ Either $$z^3=z+1\quad \Longrightarrow\quad
z\in
\F_8\setminus\F_2\subset\F_q\setminus\F_2\quad\Longrightarrow\quad
r\equiv 0 \!\!\mod  3\,,$$ and then $1+\Tr_{\F_q}
((z+1)^{-1})=\Tr_{\F_q} (z^{-1})$,

\smallskip

\noindent $\bullet$ or $1+\Tr_{\F_q} ((z+1)^{-1})=\Tr_{\F_q}
(z^{-1})=0$ and so,  $F_{q-1} (z)=0\,\,\forall z\in \F_q\setminus
\F_8$.

\smallskip

\noindent Henceforth, if $\F_q\supseteq \F_8$ and $\F_q\neq\F_8$
then $$\card (A_q\setminus \{0\})=q/2 -1 \ge q-8\quad
\Longrightarrow\quad q\le 14\quad \Longrightarrow\quad q=8\,,$$
which is a contradiction. If $\F_q\not\supseteq \F_8$ then, by
the same argument as above, $F_{q-1} (z)=0\,\,\forall z\in
\F_q\setminus \F_2$. Hence
$$\card (A_q\setminus \{0\})=q/2 -1 \ge q-2\quad
\Longrightarrow\quad q\le 2\,,$$ which again gives a
contradiction.

Therefore $\F_q=\F_8$. Indeed, for $q=2^3$ we have
$A_{q}^+=B_{q}\setminus\{0\}$ and so,  the toric lattice
$\T_{q-1,q-1}=\T_{7,7}$ is not harmonic, as stated. \eproof

\brems\label{hdivt} 1. By virtue of \ref{q-1}.b, $$r\equiv 0 \mod
2 \quad\iff\quad q-1\equiv 0 \!\!\mod 3\quad\iff\quad 0\in
A_q\cap A_q^+\cap B_q\,,$$ hence both $\T_{q-1,q-1}$ and
$\T_{q-1,q+1}$ are harmonic, and
$$r\equiv
1 \!\!\mod 2 \quad\iff\quad q+1\equiv 0 \!\!\mod 3\quad\iff\quad
0\in A_q\cap B_q\cap B_q^+\,,$$ hence both $\T_{q-1,q+1}$ and
$\T_{q+1,q+1}$ are harmonic.

2. The polynomials $h_1 (x)=x^2+x+1$ and $h_2
(x)=x^4+x+1$ satisfy $h_i(x+1)=h_i(x)$, $i=1,2$. They 
divide the Fibonacci polynomials $F_{q\pm 1}$ (and
hence also $T_{q\pm 1}$) in the following cases: $$h_1\mid F_{q-1}
\quad\iff\quad r\equiv 0\!\!\mod    4 \qquad\mbox{and}\quad
h_1\mid F_{q+1} \quad\iff\quad r\equiv 2\!\!\mod    4 \,,$$
$$h_2\mid F_{q-1} \quad\iff\quad
r\equiv 0\!\!\mod    8 \qquad\mbox{and}\quad h_2\mid F_{q+1}
\quad\iff\quad r\equiv 4\!\!\mod    8 \,.$$\erems

\bsit\label{ofi}  The above theory can be naturally extended 
to the laplacians $\D^\pm_\G$ on $\cF
(\G,\Z)$ and on $\cF (\G,\F_p)$ for all primes $p> 2$,
see e.g., \cite{MOW}, \cite{GMT}, \cite{HMP}. We say that $\G$ is {\it
$p$-(anti)harmonic} if $\ker(\D^\pm_\G)$ has a positive dimension
$d^\pm_p$ in $\cF (\G,\F_p)$. For $s$-dimensional grids, and
especially for 2-dimensional square grids, the following is
proved in \cite[\S\S 4-5]{HMP}. \esit

\bprop\label{hmp}
\begin{enumerate} \item[(a)] For $\bar
n=(n_1,\ldots,n_s)\in\N^s$, the grid 
$P_{\bar n}$  is $p$-harmonic if and only if
$$\det(\D^+_{P_{\bar n}})=\prod_{(i_1,\ldots,i_s),\,1\le i_j\le n_j}
\left( 1-\sum_{j=1}^s
\left(\varsigma_{2(n_j+1)}^{i_j}+
\varsigma_{2(n_j+1)}^{-i_j}\right)\right)\equiv
0 \!\!\mod    p\,,$$ where $\varsigma_{n}=e^{\frac{2\pi
i}{n}}\in\C$ \footnote{The above product being an integer.}.
\item[(b)] $\forall n\ge 3$ there exists a prime $p$ such that
the square grid $P_{n-1,n-1}$ and the toric lattice $\T_{n,n}$ 
are $p$-harmonic.
\item[(c)] $P_{n-1,n-1}$ ($\T_{n,n}$, respectively) 
is $p$-harmonic for every prime $p$ if and
only if $n\equiv 0  \!\!\mod   5$ or $n\equiv 0 \!\!\mod   6$ 
($n\equiv 0  \!\!\mod 5$ or 
$n\equiv 0 \!\!\mod 3$, respectively).
\item[(d)] If $l>5$ and $p$ are primes such that $p$ is a primitive
root modulo $l$ then both the square grid $P_{l-1,l-1}$ and
the toric lattice $\T_{l,l}$ 
 are not $p$-harmonic.
\item[(e)] For every prime $p$ with at most two exceptions, 
the set $I_p$ of all primes $l$ such that the square grid $P_{l-1,l-1}$ 
(the toric lattice $\T_{l,l}$, respectively)  
is not $p$-harmonic, is
infinite.
\item[(f)] The square grid $P_{n-1,n-1}$ and the toric lattice 
$\T_{n,n}$ with  $n=(p\pm 1)/2$ are $p$-harmonic for every
prime $p>23$. 
\end{enumerate}\eprop

For the proof of (a), (b), (d), (f) see 
\cite{HMP}, 4.6, 4.4, 4.3, 4.7 and 5.4, respectively.  
The proof of (e) in \cite{HMP} is based on (d) and on a result of Heath-Brown
\cite{HB} on Artin's conjecture of primitive roots. 
\footnote{This conjecture suggests that every integer $n\neq -1$ 
which is not a square, 
is a
primitive root modulo $l$ for an infinite set, say, $I_n$ of primes $l$. 
The 
result of Heath-Brown {\it loc.cit.} says that the property in 
Artin's conjecture holds
for all primes
with at most 2 exceptions, and for all square-free integers with at
most 3 exceptions. However, so far no concrete example of a 
prime satisfying the conjecture
has been found, see \cite{HB, Mo, Mu}.} 

\bsit\label{dplus} For a graph $\G$ we let as before $d^\pm
(\G)=\dim (\ker (\D^\pm_\G))=\dim (\Harm^\pm (\G))$. The kernels
of the grid laplacians $\D^\pm_{P_{m,n}}$ admit the following
description, see \cite{Su2}, \cite[3.6, 3.10]{Su3}, \cite[3.6, 4.1]{BR}
and \ref{phe} below.\esit

\bprop\label{kern}
\begin{enumerate}\item[(a)] $\forall p\in k[x]$, $\dim
(\ker (p(\D^-_{P_m})))=\deg (\gcd (p, E_m))$.
\item[(b)] $\ker (\D^\pm_{P_{m,n}})\cong \ker (E_m(\D^\pm_{P_{n}}))\cong
\ker (E_n(\D^\pm_{P_{m}}))$.
\item[(c)] Furthermore, $\ker (\D^-_{P_{m-1,n-1}})\cong \ker
(E_{\gcd (m,n)-1} (\D^-_{P_{n-1}}))$.
\item[(d)] Consequently, $d^-(P_{m-1,n-1})=\gcd (m,n)-1$ and
$d^+(P_{m-1,n-1})=\deg (\gcd(E_{m-1}, E_{n-1}^+))$.
\item[(e)] $\forall m=2^r$, $\forall n=2^kp$, where $r\ge 1$ 
and $p\in\NO$,
one has \footnote{See also \cite[5.2]{Su3} for the case $m=3\cdot
2^r$, $n=2^kp$.} $$d^+(P_{m-1,n-1})=d^+ (P_{n-1})=\begin{cases} 0
& \mbox{if}\quad p\not\equiv 0 \!\!\!\mod  3,\,\, (\mbox{and
so}\,\, P_{m-1,n-1}
\,\,\mbox{is not harmonic}), \\
2^{k+1} & \mbox{if}\quad p\equiv 0 \!\!\!\mod  3\quad\mbox{and}
\quad k<r-1,\\
m-1 &\mbox {otherwise}\,.
\end{cases}$$ \item[(f)] Moreover $\min\{n\ge m\,:\, d^+(P_{m-1,n})=m-1
\}=\frac{3}{2}m-1$.
\end{enumerate}\eprop

\bexas\label{exa0} The path $P_2$ and the grids $P_{2,2n-1}$,
$n\ge 1$, $P_2\times C_n$ and $\T_{3,n}$, $n\ge 3$, are harmonic,
whereas $P_{2,2n}$, $n\ge 1$, are not. The grid $P_{2,3}$ has the
nuclei
$$ \begin{pmatrix}
  1 & 1 & 1 \\
  0 & 1 & 0
\end{pmatrix},\qquad \begin{pmatrix}
  0 & 1 & 0 \\
  1 & 1 & 1
\end{pmatrix} \quad\mbox{and}\quad \begin{pmatrix}
  1 & 0 & 1 \\
  1 & 0 & 1
\end{pmatrix}\,.
$$ Thus $d^+ (P_{2,3})=2$. Similarly,
the grid $P_{2,2n-1}$ ($P_{2,4n-1}$, respectively) has a nucleus
$$\begin{pmatrix}
  1 & 0 & 1 & 0 & 1 &  0 & \ldots & 1 \\
  1 & 0 & 1 & 0 & 1 &  0 & \ldots & 1
\end{pmatrix},\quad\mbox{respectively,}\quad
\begin{pmatrix}
  0 & 1 & 0 & 0 & 0 & 1 & 0 & 0 & \ldots & 0 \\
  1 & 1 & 1 & 0 & 1 & 1 & 1 & 0 & \ldots & 1
\end{pmatrix}\,.$$ \eexas

\subsection{$n$-dimensional case} From \ref{bache} we deduce by
recursion the following, cf. \cite{SB}.

\bprop\label{prod1}
\begin{enumerate}\item[(a)] $\spec^- (\prod_{i=1}^s
\G_i)=\sum_{i=1}^s\spec^- (\G_i)$ and $\spec^+ (\prod_{i=1}^s
\G_i)=1+\sum_{i=1}^s\spec^- (\G_i)$.

\item[(b)] For the product $\G=\prod_{i=1}^s \G_i$ of $s_1$ 
harmonic and
$s-s_1$ antiharmonic graphs we have $(s_1+1)\!\!\!\mod  2
\in\spec^+ (\G)$. Consequently, $\G$ is harmonic if $s_1$ is odd
and $\G$ is antiharmonic otherwise. If at least one of the factors
$\G_i$ is biharmonic then so is $\G$.

\item[(c)] $\forall f_i\in\Harm^+_{\bar k}
(\G_i),\,i=1,\ldots,s_1$, $\forall g_j\in\Harm^-_{\bar k}
(\G_j),\,j=s_1+1,\ldots,s$, the function
$h=(\bigotimes_{i=1}^{s_1} f_i)\otimes (\bigotimes_{j=s_1+1}^{s}
g_j)\in \cF (\G, \bar k)$ is harmonic for $s_1$ odd and
antiharmonic for $s_1$ even.

\item[(d)]  If $N_i^+$ is a nucleus of $\G_i$, $i=1,\ldots,s_1$,
and $N_j^-$ is an antinucleus of $\G_j$, $j=s_1+1,\ldots,s$, then
$N=\prod_{i=1}^{s_1} N_i^+\times \prod_{j=s_1+1}^s N_j^-$ is a
nucleus of $\G=\prod_{i=1}^s \G_i$ for $s_1$ odd and an
antinucleus of $\G$ for $s_1$ even.\end{enumerate}\eprop

\bsit\label{ovrl} We keep 
the notation 
$$\T_{\bar n}=\prod_{i=1}^s
C_{n_i}, \qquad P_{\bar n}=\prod_{i=1}^s P_{n_i}\quad\mbox{ and}\quad 
\overline
{n-1}=(n_1-1,\ldots,n_s-1)\,.$$ For the next results see e.g. 
\cite{Su1}, \cite[\S 5-6]{SB}, 
\cite[\S 3]{HMP}). \esit

\bprop\label{prod2} \begin{enumerate}
\item[(a)]  For any graph $\G$ and for every $n\in\N$,
$$\chi_{\G\times P_{n-1}}^{\ } (x)= \Res_y \left(\chi_{\G}^{\ }(x+y),
F_{n}(y)\right), \qquad \chi_{\G\times C_{n-1}}^{\ } (x)= \Res_y
\left(\chi_{\G}^{\ }(x+y), T_{n}(y)\right)\,$$ and
$$\spec^\pm (\G\times C_n)=\{\lambda\in\bar k^\times\,:\, 
\lambda+\lambda^{-1}\in
\spec^\pm (\G)\}\,.$$ Hence $\G\times C_n$ is harmonic if and only
if $1+\lambda+\lambda^{-1}\in \spec^+ (\G)$ for some
$\lambda\in\mu_n$.
\item[(b)]
We have
$$\spec^+ (\T_{\bar n})=\{1 + \sum_{i=1}^s (\xi_i+\xi_i^{-1})\,:\,
\xi_i\in \mu_{n_i},\,\, i=1,\ldots,s\}\,.$$ Thus $\T_{\bar n}$ is
harmonic if and only if the system (\ref{ME}) in \ref{resum} has a
solution $(x_1,\ldots,x_s)\in ({\bar k}^\times)^s$.
\item[(c)] If $\G\times C_n$ is harmonic then so is $\G\times C_{ln}$ 
for
every $l\ge 1$.
\item[(d)] If $\T_{\bar n}$ is harmonic then so
is $\T_{\bar n'}\times \T_{\bar m}$, $\forall \bar m\in\N^t$,
$\forall \bar n'=(l_1n_1,\ldots,l_sn_s)\in\N^s$, where $l_i\ge
1\,\forall i=1,\ldots,s$.
\item[(e)] If $\G\times C_{2n}$ ($n\ge 3$) is harmonic then so is 
$\G\times
C_n$. Consequently,  $\G\times C_{2^r}$ ($r\ge 2$) is harmonic if
and only if so is $\G$.
\item[(f)] $\forall \bar n=(2^{r_1},\ldots,2^{r_s})$, the toric lattice
$\T_{\bar n}$ is not harmonic. \end{enumerate} \eprop

\bproof (a) follows from \ref{charpo} and \ref{bache}, and implies
(b) by recursion. The covering $\G\times C_{ln}\to \G\times C_n$
with the Galois group $\Z/l\Z$ induces the injections $$\pi^* :
\Harm^\pm (\G\times C_n) \hookrightarrow \Harm^\pm (\G\times
C_{ln})\,, \qquad f\longmapsto f\circ\pi\,.$$ This proves the
harmonicity of $\T_{\bar n'}$ in (d), whereas that of the product
$\T_{\bar m}\times \T_{\bar n}$ follows from \ref{prod1}.b. The
proof of (e) uses (a) and the fact that
$\phi: \bar k\to \bar k,\,x\longmapsto x^2$ is an automorphism. (f)
follows from (e) by recursion. \eproof

\bexas\label{exa}  For any antiharmonic graph $\G$ and $\forall
n\ge 3,\,\forall l\ge 1$, the products $\G\times C_n$ and
$\G\times P_{2l-1}$ are antiharmonic , whereas $\G\times C_{3l}$
and $\G\times P_{3l-1}$ are harmonic. If $\G$ is harmonic then so
are the products $\G\times C_n$ and $\G\times P_{2l-1}$, whereas
$\G\times C_{3l}$ and $\G\times P_{3l-1}$ are antiharmonic. See
also \cite[\S\S 5,6]{SB} for the (anti)harmonicity of the
hypercubic grids $P_{\bar n}$ and of the products $P_{\bar n}\times \T_{\bar
m}$.
 \eexas

\subsection{Symmetrization}
\bsit\label{symme} Let $K$ be a field with char$(K)=2$. If
$\alpha: \G\to \G$ is an involution then for any nonzero $f\in
\cF (\G,K)$, either $f\circ\alpha=f$ or the average
$g=f+f\circ\alpha$ is again nonzero and is $\alpha$-stable:
$g\circ\alpha=g$. Anyhow, if $\Harm^\pm_K (\G)\neq \{ 0\}$ then
also  $[\Harm^\pm_K (\G)]^{\alpha}\neq \{ 0\}$. Moreover, if
$F=\Fix (\alpha)\neq\emptyset$ then for any $f\in [\cF
(\G,K)]^{\alpha}$, \be\label{restr}\D_\G^\pm(f)\vert F=
\D^\pm_{F}(f\vert F)\,,\ee
  and so the restriction of an
$\alpha$-symmetric (anti)harmonic function $f$ to $F$ is again
(anti)harmonic. Furthermore, if $f\vert F\equiv  0$ then also
$f\vert (\G\ominus F)\in \Harm^\pm (\G\ominus F)$. Thus if $\G$
is (anti)harmonic then so is at least one of the graphs $F$ and
$\G\ominus F$. In case that $F$ is a 'separation wall` for $\G$,
from the above discussion we deduce the following result.\esit

\blem\label{sym2} Let $\alpha$ be an involution of
$\G$ such that $F={\rm Fix} (\alpha)$ separates $\G$
that is, $\G\ominus F=\G^+\cup \G^-$, where $\G^+$ and $\G^-$ are
two disjoint subgraphs of $\G$ with $\alpha
(\G^\pm)=\G^\mp$. If $\G$ is (anti)harmonic then so is at least
one of the graphs $F$ and $\G^\pm$. \elem

\bcor\label{sym3}  \begin{enumerate}\item[(a)] If $\G\times
P_{n-1}$ is (anti)harmonic then so is $\G\times C_{n}$. Vice
versa, if $\G\times C_{n}$ is (anti)harmonic then so is at least
one of the graphs $\G$ and $\G\times P_{n-1}$.

\item[(b)] (cf. \cite[6.1]{SB})
Consequently, if the grid $P_{\overline {n-1}}= \prod_{i=1}^s
P_{n_i-1}$ is harmonic, where $\bar {n}=(n_1,\ldots,n_s)$ 
then so is the toric lattice $\T_{\bar
{n}}$. In particular for
every $\bar n=(2^{r_1}-1,\ldots,2^{r_s}-1)$ the grid $P_{\bar n}$
is not harmonic.

Vice versa, if $\T_{\bar {n}}$ is harmonic and $\prod_{i=1}^s
n_i\not\equiv 0 \!\!\!\mod  3$ then $P_{\overline {n-1}}$ is
harmonic too.

\item[(c)] If $\G\times P_{n-1}$ is harmonic then so is $\G\times
P_{ln-1}$ $\forall l\ge 1$.

\item[(d)] If $\G\times P_{2n+1}$ is (anti)harmonic then so is at least one
of the graphs $\G$ and $\G\times P_n$.\end{enumerate}\ecor

\bproof (a) follows from \ref{sym2}. To show (b), letting $f$ be
a nonzero (anti)harmonic function on $\G\times P_{n-1}$ symmetric
w.r.t. the reflection
$$\alpha: P_{n-1}\to P_{n-1},\quad
v_i\longmapsto v_{(n-1-i)\!\!\mod (n-1)}\,,$$ the extension of
$f$ by zero to $\G\times C_{n}\supseteq \G\times P_{n-1}$ is again
(anti)harmonic, as required. The converse in (b) follows from
\ref{sym2}. Iterating this argument yields the first and the last
assertions of (c). The second one follows by \ref{prod2}.d.

To show (d) we take $l$ copies $P_{n-1}^{(i)}$, $i=1,\ldots,l$ of
$P_{n-1}$. For a nucleus $N$ of $\G\times P_{n-1}$, we consider
its copy $N_1$ in $\G\times P_{n-1}^{(1)}$, the mirror image
$N_2$ of $N_1$ in $\G\times P_{n-1}^{(2)}$, the mirror image
$N_3$ of $N_2$ in $\G\times P_{n-1}^{(3)}$, etc. Taking also new
vertices $v_1',\ldots,v_{l-1}'$ and representing $\G\times
P_{ln-1}$ as ordered 'connected sum` of the graphs
$$\G\times P_{n-1}^{(1)},\quad \G\times \{v_1'\},
\quad \G\times P_{n-1}^{(2)},\quad \G\times \{v_2'\},\quad
\ldots,\quad\G\times \{v_{l-1}'\}, \quad \G\times P_{n-1}^{(l)}$$
we obtain a nucleus $N'=\bigcup_{i=1}^l N_i$ of $\G\times
P_{ln-1}$. \eproof

\brems\label{sym4} 1. Starting with $N=P_2$ the proof of (d)
gives a nucleus of $P_{3l-1}$ $\forall l\ge 1$.

2. Instead of taking average of $f$ over an involution, one can
consider the average of $f$ over the shifts on $\Z_{\bar n}$.
Suppose that for some $i\in\{1,\ldots,s\}$, the subgroup $\Z_{\bar
n^{(i)}}$ of $\Z_{\bar n}$ is not harmonic, where $\bar
n^{(i)}=(n_1,\ldots,n_{i-1},n_{i+1},\ldots,n_s)$. Then for any
$f\in\Harm^+ (\Z_{\bar n})$, the average $f+f_{\vec
e_i}+\ldots+f_{(n_i-1)\vec e_i}$ of $f$ over the shifts by the
subgroup $\Z_{n_i}\subseteq \Z_{\bar n}$ must be zero. This means
that the intersection of each nucleus $N^+$ of $\Z_{\bar n}$ with
every 'line` $l_{p,i}=\{p, p+\vec e_i,\ldots,p+(n_i-1)\vec
e_i\}$, $p\in \Z_{\bar n}$, has even cardinality. \erems

\bsit\label{caley} For a vertex $v$ of a graph $\G$ we let
\be\label{al} a_v^+=\delta_v+\sum_{[u,v]\in\EG}
\delta_u\in\cF(\G, k)\,.\ee Actually the toric lattice
$\T_{\bar n}$ represents the Caley graph of the group $\Z_{\bar
n}=\prod_{i=1}^s \Z/n_i\Z$ with its standard generators
$(e_i)_{i=1,\ldots,s}$. Every involution $\alpha'$ of $\T_{\bar
n}$ with a fixed point is conjugated with an involutive
automorphism $\alpha : \Z_{\bar n}\to \Z_{\bar n}$ stabilizing
$a_e^+$: $a^+_e\circ\alpha= a^+_e$. Moreover $\alpha (e_i)=\pm
e_{\sigma (i)}$, where $\sigma\in S_s$ is a product of independent
transpositions such that $n_i=n_{\sigma (i)}\,\,\forall
i=1,\ldots,s$. The induced action of $\alpha$ on $\cF (\T_{\bar
n}, K)$ commutes with $\Delta=\Delta^+_{\T_{\bar n}}$:
$$(\Delta f)\circ\alpha=\Delta (f\circ\alpha)\quad \forall f\in
\cF (\T_{\bar n}, K)\,.$$ Hence $\Delta^k(\delta_e)\vert F =
\Delta^k_F (\delta_e\vert F)$, where $F=\Fix (\alpha)$. In
particular, if $\Z_{\bar n}$ is not harmonic then so is $F$.
Choosing $\alpha$ appropriately, we arrive at the same conclusion
as in \ref{prod2}.d.\esit

\subsection{Uniqueness sets} Let $K$ be a field of characteristic 2.
A subset $U\subseteq \VG$ is called a {\it uniqueness set} for
$\Harm^\pm_K (\G)$ if every function $f\in \Harm^\pm_K (\G)$ that
vanishes on $U$ vanishes identically. Thus every (anti)harmonic
function $f$ on $\G$ is uniquely determined by its restriction
$f\vert U$.

The boundary of a bounded plane domain
is a  uniqueness set for the classical harmonic functions. 
In our discrete
setting, it may happen that just a part of the boundary (or of the
interior) serves as a uniqueness set for binary harmonic functions. 
Let us give several examples.

\bexas \label{ue} 1. An extremal vertex of the linear string $P_n$
is a uniqueness set for $\Harm^\pm_K (P_n)$. Every pair of
neighborhooding vertices of the circular graph $C_n$ is a
uniqueness set for $\Harm^\pm_K (C_n)$.

2. More generally, $\G\times \{v_1\}$ and $\G\times
\{v_i,v_{i+1}\}$, $2\le i\le n-2$, are uniqueness sets for the
(anti)harmonic functions on $\G\times P_n$, and $\G\times
\{v_i,v_{i+1}\}$ is that on $\G\times C_n$.

3. The set of all extremal vertices of a finite forest $\G$ is a
uniqueness set for the (anti)harmonic functions on $\G$. On the
other hand, the reduction $\G^+_{\rm red}$ ($\G^-_{\rm red}$,
respectively) as in \ref{redu} regarded as a subgraph of $\G$ is
a uniqueness set for harmonic (respectively, antiharmonic)
functions on $\G$, see the proof of \ref{fore}.

4. Every side of a triangle $\Pi_n$ inscribed in a triangular
plane lattice is a uniqueness set for $\Harm^\pm_K (\Pi_n)$.

5. The exterior circle is a uniqueness set for the conic lattice
$C_n (m)$ made of $m$ concentric plane copies of $C_n$ joint one
to another by radial edges, the last copy being also joint with a
new vertex at their common center.
 \eexas

\subsection{Periodic harmonic extension}
The idea behind \ref{prod2}.a and \ref{kern} is as follows, cf.
\cite{Su3, BR}.

\bsit\label{phe} To any function $f\in \cF (\G\times C_n, K)$ one
associates a sequence $f_1,\ldots,f_n\in \cF (\G, K)$, where
$f_i=f\vert (\G\times \{v_i\})$. Letting $\D^\pm=\D^\pm_{\G\times
C_n}$ we obtain
$$(\D^\pm f)_i=f_{(i-1)\!\!\!\mod n}+
\D^\pm_\G (f_i)+f_{(i+1)\!\!\!\mod n},\qquad i
=1,\ldots,n\,.$$ Therefore $f\in \Harm^\pm (\G\times C_n)$ if and
only if \be\label{peri} f_{(i+1)\!\!\!\mod n}=f_{(i-1)\!\!\!\mod
n}+\D^\pm_\G (f_i)\qquad\forall i=1,\ldots,n\,.\ee Starting with
an arbitrary pair $u_0=(f_0,f_1)\in V:=[\veg]^2$ and applying
successively the automorphism
$J_\G=J_\G^\pm=\left(\begin{array}{cc} 0& 1
\\ 1 & \D^\pm_\G\end{array}\right)\in\Aut (V)$ we extend $u_0$ to
a function $f$ on $\G\times C_n$ so that
$$(f_1,f_2)=u_1=J_\G(u_0)=J_\G(f_0,f_1),\ldots, (f_n,f_{n+1})=u_n
=J_\G(u_{n-1}) =J_\G(f_{n-1},f_n)\,.$$ This extension $f$ is
(anti)harmonic provided that it is periodic. The latter holds if
and only if $J_\G^n (u_0)=u_0$. By recursion we obtain
$$J_\G^n=\begin{pmatrix}
  F_{n-1}(\D^\pm_\G) & F_{n}(\D^\pm_\G)  \\
  F_{n}(\D^\pm_\G)  & F_{n+1}(\D^\pm_\G)
\end{pmatrix}\,.$$ Thus $$J_\G^n
(u_0)=u_0\quad\Longleftrightarrow\quad
\begin{cases} F_{n-1}(\D^\pm_\G)f_0+F_{n}(\D^\pm_\G)f_1=f_0\\
F_{n}(\D^\pm_\G)f_0+F_{n+1}(\D^\pm_\G)f_1=f_1\,.
\end{cases}$$
In particular $(0,f_1)\in \ker (\id+J_\G^n)\quad\iff\quad
f_1\in\ker (F_{n}(\D^\pm_\G))\cap \ker (\id+F_{n-1}(\D^\pm_\G))$.

Hence $\G\times C_n$ is (anti)harmonic if and only if $1 \in {\rm
spec} (J_\G^n)$, or equivalently, if there exists $\lambda \in
{\rm spec} (J_\G)\cap \mu_n$. We have
$${\rm spec} (J_\G^\pm)=\{\lambda\in \bar k^\times\,:\,
\lambda+\lambda^{-1}\in\sigma^\pm_\G\}\,.$$ Thus $\G\times C_n$
is (anti)harmonic if and only if there is $\lambda \in \mu_n$ such
that $\lambda+\lambda^{-1}\in\sigma^\pm_\G$. This proves the
second assertion in \ref{prod2}.a.\esit

\brem\label{symext} If $\G\times C_n$ is harmonic then, according
to the symmetrization and uniqueness principles, every
(anti)harmonic function on $\G\times C_n$ with $f_0=f_1$ is
necessarily symmetric i.e., $f_i=f_{(-i+1)\!\!\mod   n}$ for all
$i=1,\ldots,n$.\erem

We let $m=\card (\VG)$. Since $0\notin \spec (J_\G^\pm)$,
$J_\G^\pm\in\GL_k (2m)$ has finite order i.e., $\exists
n\in\N\,:\, J^n=1\quad\iff\quad F_{n-1}
(\D^\pm_\G)+\id=0=F_n(\D^\pm_\G)$. The previous discussion leads to
the following result.

\bprop\label{newperi} $\forall n\ge 3$, $d^\pm (\G\times C_n)\le
\dim_k (V)=2\card (\VG)$. The equality holds if and only if
$n\equiv 0 \mod\,   \ord\,J_\G^\pm$. \eprop

\bexas\label{3,6} 1. For $\G=C_3$, $\ord (\D^+_\G)=2$ and $\ord
(J_\G^+)=6$. So $d^+(\T_{3,n})\le 6$ $\forall n\ge 1$ and
$d^+(\T_{3,n})=6\quad\iff\quad n\equiv 0\!\!\mod   6$. The cyclic
shifts in the vertical direction of the harmonic patterns 
$$h_1=\begin{pmatrix}
  1 & 0 & 1 & 1 & 0 & 1 \\
  0 & 0 & 0 & 1 & 1 & 1 \\
  0 & 0 & 0 & 1 & 1 & 1
\end{pmatrix},\quad h_2=\begin{pmatrix}
  0 & 1 & 1 & 0 & 1 & 1 \\
  0 & 0 & 1 & 1 & 1 & 0 \\
  0 & 0 & 1 & 1 & 1 & 0
\end{pmatrix}$$ form a basis of $\Harm^+ (\T_{3,6})$.
The 2-sheeted covering $\pi:\T_{3,6} \to \T_{3,3}$ yields 
4-dimensional subspace $\pi^*(\Harm^+ (\T_{3,3}))\subseteq \Harm^+
(\T_{3,6})$.

2. Likewise, for $\G=C_5$, $\ord (\D^+_\G)=3$ and $\ord
(J_\G^+)=15$, so $d^+(\T_{5,n})\le 10$ $\forall n\ge 1$ and
$d^+(\T_{5,n})=10\quad\iff\quad n\equiv 0\!\!\mod   15$. The
3-sheeted covering $\pi:\T_{5,15} \to \T_{5,5}$ gives rise to 
8-dimensional subspace $\pi^*(\Harm^+ (\T_{5,5}))\subseteq \Harm^+
(\T_{5,15})$.\eexas

\bsit\label{perpat} Similarly, for the path $P_{n-1}$, the
function $f=(f_1,\ldots,f_{n-1})\in \cF(\G\times P_{n-1},K)$ is
(anti)harmonic if and only if (\ref{peri}) holds for all
$i=1,\ldots,n-1$ with $f_0=f_n=0$. By recursion, the latter holds
if and only if $f_k=F_k(\D_\G^\pm)f_1$ $\forall k=2,\ldots,n$.
Thus $f_1\in\cF(\G,K)$ extends to $f\in \Harm^\pm (\G\times
P_{n-1},K)$ if and only if $F_{n+1} (\D_\G^\pm)(f_1)=0$. Hence
$\Harm^\pm (\G\times P_{n-1},K)\cong \ker (E_n(\D_\G^\pm))$. This
shows \ref{kern}.b.\esit

\subsection{Doubling the periods}

\bsit\label{schar1} In this subsection we consider an 
 injective $k$-endomorphism $\delta: \cF(\Z^2,k)\to 
 \cF(\Z^2,k)$, which sends a function
 $$f=\begin{pmatrix}
  \cdots & \cdots & \cdots & \cdots & \cdots \\
  \cdots & r & s & t & \cdots \\
  \cdots & u & v & w & \cdots \\
  \cdots & x & y & z & \cdots \\
  \cdots & \cdots & \cdots & \cdots & \cdots
\end{pmatrix}
$$ into 
 $$
 \delta (f)=\begin{pmatrix}
  \cdots & \cdots & \cdots & \cdots & \cdots & \cdots & \cdots\\
  \cdots & r & r+s & s & s+t & t & \cdots \\
  \cdots & r+u & 0 & s+v & 0 & t+w & \cdots \\
  \cdots & u & u+v & v & v+w & w & \cdots \\
  \cdots & u+x & 0 & v+y & 0 & w+z & \cdots \\
  \cdots & x & x+y & y & y+z & z & \cdots \\
  \cdots & \cdots & \cdots & \cdots & \cdots & \cdots & \cdots
\end{pmatrix}\,.
$$ For instance, $\delta$  sends the harmonic function 
$$h_0= \begin{pmatrix}
  \cdots & \cdots & \cdots & \cdots & \cdots & \cdots & \cdots\\
  \cdots & 1 & 1 & 1 & 1 & 1 & \cdots \\
  \cdots & 1 & 1 & 1 & 1 & 1 & \cdots \\
  \cdots & 0 & 0 & 0 & 0 & 0 & \cdots \\
  \cdots & 1 & 1 & 1 & 1 & 1 & \cdots \\
  \cdots & 1 & 1 & 1 & 1 & 1 & \cdots \\
  \cdots & \cdots & \cdots & \cdots & \cdots & \cdots & \cdots
\end{pmatrix}
$$ to the harmonic function 
\footnote{Respectively, $\,^{t}h_0$ to $\delta(\,^{t}h_0)=\,^{t}\delta(h_0)$.}
$$\delta (h_0)= \begin{pmatrix}
  \cdots & \cdots & \cdots & \cdots & \cdots & \cdots & \cdots\\
  \cdots & 1 & 0 & 1 & 0 & 1 & \cdots \\
  \cdots & 1 & 0 & 1 & 0 & 1 & \cdots \\
  \cdots & 0 & 0 & 0 & 0 & 0 & \cdots \\
  \cdots & 1 & 0 & 1 & 0 & 1 & \cdots \\
  \cdots & 1 & 0 & 1 & 0 & 1 & \cdots \\
  \cdots & \cdots & \cdots & \cdots & \cdots & \cdots & \cdots
\end{pmatrix}
$$  
replacing the bi-period $\bar n (h_0)=(1,3)$ 
by the bi-period $\bar n (\delta (h_0))=(2,3)$.\esit

 \bprop\label{schar} 
 The endomorphism $\delta$ 
 stabilizes the subspace $\Harm^+ (\Z^2)$ and generically
doubles bi-periods. More precisely, if $f\in \Harm^+ (\Z^2)$ 
is a bi-periodic binary harmonic function with
bi-period $\bar n (f)=(n_1,n_2)$ different from a shift of
$h_0$ or $\,^{t}h_0$, then $n_1,n_2>1$ and
$\delta (f)\in \Harm^+ (\Z^2)$ has bi-period 
$\bar n (\delta (f))=2\bar n (f)=(2n_1,2n_2)$. 
\eprop

\bproof It is easily seen that $\delta (f)$ is harmonic if so is $f$. 
If $f\neq 0$ is constant in vertical or horizontal direction 
then it is a shift of one of the functions $h_0$ or $\,^{t}h_0$, 
which has been excluded.
Thus $n_1, n_2>1$, the function $f$ on
  $\Z^2$ is non-constant in vertical direction, and so
  $\delta (f)$ possesses a nonzero 
  line $$\begin{pmatrix}
  \cdots & u+x & 0 & v+y & 0 & w+z & \cdots \\
\end{pmatrix}\,.$$ If
  $(m_1,0)$ is a  period of $\delta (f)$ then necessarily 
  $m_1$ is even and
  $(m_1/2,0)$ is a  period of $f$, so $(m_1,0)=(2n_1,0)$ 
  is a minimal such period. 
  By symmetry $\bar n (\delta (f))=(2n_1,2n_2)$, as stated. 
  \eproof
  
   \brems\label{double} 1. Actually $\delta$ provides linear injections 
   $\Harm^+ (\T_{n_1,n_2})\hookrightarrow\Harm^+ (\T_{2n_1,2n_2})$.
   
   2. Doubling of just one of
the periods $n_1,n_2$ is impossible in general, 
as Example \ref{3,6}.1 above shows.
   
   3. A similar doubling is equally 
   applied in higher dimensions, over any 
   field $K$ of characteristic 2. 
   Namely, for any $f\in \cF(\Z^s,K)$ and $u\in \Z^s$
   we let 
   $\delta(f) (u)=f(v)$ if $u=2v$ has
   all coordinates even, otherwise 
   $$\delta(f) (u)=\sum_{2v\in {\rm neighb} (u)} f (v)\,,$$ 
   where $2v\in {\rm neighb} (u)$ iff $v\in\Z^s$ and
   the coordinates of $u-2v$ are equal to $0$ or $\pm 1$.  \erems

\subsection{'Lights Out` game on graphs}

The game 'Lights Out` on a finite graph $\G$ consists in the
following  \cite{Pe}, \cite{Su4}. Each vertex of $\G$ 
can be in one of the
two states 'on` or 'off`. A move consists in changing the state
of a vertex and, simultaneously, of all its neighbors. The goal
is to get, after a sequence of moves, all states 'off`. An
initial position will be called {\it pattern}. A pattern is {\it
winning} if there exists a sequence of moves terminating at a
zero pattern. The graph $\G$ is called {\it winning} if the game
on $\G$ wins starting with an arbitrary pattern. Most of the
following results are well known, see e.g., \cite{Su5}.

\bprop\label{game} \begin{enumerate}\item[(a)] A finite graph $\G$
is winning if and only if $\G$ is not harmonic.
\item[(b)] Every nucleus $N^+$ of $\G$ yields a linear relation
for the functions $\{a_v^+\}_{v\in\VG}$: $$\D^+_\G
(\delta_{N^+})=\sum_{v\in N^+} a_v^+= 0\,,$$ where
$\delta_N$ is the characteristic function of $N$ and $a_v^+$ is
as in (\ref{al}).
\item[(c)] The space of winning patterns
$V_\G={\rm span}\, (a_v^+ \,:\,  v\in\VG) $ in $\cF (\G, k)$
is the orthogonal
complement to $\Harm^+ (\G)$ w.r.t. the standard
bilinear form $\langle x,y\rangle$ on $\cF (\G, k)$.
\item[(d)] Every antiharmonic pattern on $\G$ is winning.
\item[(e)] For every $\G$, the all-on pattern is winning.
\end{enumerate}\eprop

\bproof (a) A pattern can be considered as a binary function
on $\G$. The move at a vertex $v$ corresponds to the shift  by
$a_v^+$ in $\cF(\G, k)$. Thus the game on $\G$ is winning if and
only if the group of translations generated by
$\left(t_{a_v^+}\,:\, v\in\VG\right)$ acts transitively on
$\cF(\G, k)$. The latter holds if and only if the functions
$a_v^+$, where $v\in\VG$, span $\cF(\G, k)$, if and only if the
matrix $I+\adj (\G)$ of $\D_\G^+$ with columns $(a_v^+)$ is
non-degenerate, or, equivalently, $ 0\notin \spec^+ (\G)$, as
stated.

By definition, $\delta_N\in\Harm^+ (\G)$ if and only if $N$ is a
nucleus of $\G$. This yields (b). Now (c) follows from (b). The
proper subspaces $\Harm^+ (\G)$ and $\Harm^- (\G)$ of the
laplacian $\D^-_\G$ being orthogonal, (d) follows by virtue of
(c). Further by (c), $1\in V_\G$ if and only if $1\bot\, \Harm^+
(\G)$. The latter holds indeed because for any $h\in \Harm^+
(\G)$, the nucleus $h^{-1} (1)$ being an odd graph, by the
handshaking theorem it has an even number of vertices. \eproof

\brems\label{GE} 1. In view of \ref{game}.c the harmonic functions
on $\G$ are linear invariants of the game 'Lights Out`. That is,
$$\langle h, f+ a_v^+\rangle=\langle h, f\rangle\qquad
\forall v\in\VG,\,\forall f\in \cF(\G, k),\,\forall h\in \Harm^+
(\G)\,.$$ By virtue of \ref{game}.a, $\G$ is winning if and only
if it does not admit a nonzero linear invariant.

2. \ref{game}.e is Sutner's Garden-of-Eden theorem \cite{Su5}.
The desired transformation of the all-one pattern into the
all-zero one is achieved via moves at the vertices of an {\it
odd-domination subgraph} $N$ of $\G$. The latter means that every vertex $v$
of $\G$ must have in $N$ an odd number of neighbors including
$v$ itself if $v\in N$, so that $1=\sum_{v\in N} a_v^+\in V_\G$.
Given any graph
$\G$, Sutner's theorem actually proves the existence 
of an odd-domination subgraph of $\G$ 
(see also \cite{An}). \erems

\section{Counting harmonic toric 2-lattices and counting points on an
elliptic cubic curve}

\subsection{Constructing harmonic tori from polynomials}

\bprop\label{topo} Given a polynomial $p(x)\in k[x]$ with $p(0)=
1$ and a root $z\in\F_q$ of $p$, one can construct a harmonic toric
lattice $\T_{\bar n}$, and every such lattice appears that way.
\eprop

\bproof Indeed, $z$ can be written in a unique way as
$z=\zeta+\zeta^{-1}$, where $\zeta\in\F_{q^2}$. There is a unique
decomposition
$$p= 1+\sum_{0<\alpha_1<\alpha_2<\ldots<\alpha_s} T_{\alpha_i}\,,$$
and $p(z)= 0$ yields
$$\sum_{0<\alpha_1<\alpha_2<\ldots<\alpha_s}
T_{\alpha_i}(\zeta+\zeta^{-1})
=\sum_{0<\alpha_1<\alpha_2<\ldots<\alpha_s}
(\zeta^{\alpha_i}+\zeta^{-\alpha_i}) = 1\,.$$ Thus letting
$x_i=\zeta^{\alpha_i}$ gives a solution of  (\ref{ME}) with
$$n_i=\ord\, x_i = \frac{n}{\gcd (n, \alpha_i)}\,,\qquad
i=1,\ldots,s\,,$$ where $n=\ord\, \zeta=\ford\,z$.

In this way we obtain all harmonic toric lattices $\T_{\bar n}$. 
Indeed, given $\bar n=(n_1,\ldots,n_s)$ and
$m=\lcm (n_1,\ldots,n_s)$, we let $q=2^{f(m)}$, and we fix a
primitive $(q-1)$-st root of unity $\zeta\in\mu_{q-1}$. Any
solution $\bar x=(x_1,\ldots,x_s) \in(\F_q^\times)^s$ of
(\ref{ME}) can be written as $x_i=\zeta^{\alpha_i}$, where
$\alpha_in_i\equiv 0\!\!\mod   (q-1)$, $i=1,\ldots,s$. Letting
$z=\zeta+\zeta^{-1}\in\F_q$, the first equation in (\ref{ME}) is
equivalent to $p(z)=0$, where $p=1+\sum_{i=1}^s T_{\alpha_i} \in
k[x]$. \eproof

\brem\label{manypr} Letting above $\zeta=\xi^c$, where $\gcd
(c,q-1)=1$, we obtain $z=T_c(z')$ and $x_i=\xi^{c\alpha_i}$, where
$z'=\xi+\xi^{-1}\in\F_q$ is a root of the polynomial $p_c=1+
\sum_{i=1}^s T_{c\alpha_i}$.\erem

 \bexa\label{nn} For every $n=2s+1$ odd and for every $s$-tuple
$\bar n= (n,\ldots,n)$, the hypercubic toric lattice $\T_{\bar n}$
is harmonic. Indeed, if $\zeta\in\bar k$ is a primitive $n$-th root
of unity then $x_j=\zeta^j$, $j=1,\ldots,s$, gives a solution of
(\ref{ME}) with $n_j=n\,\forall j$. In particular $\T_{(5,5)}$,
$\T_{(7,7,7)}$, $\T_{(11,11,11,11,11)}$ etc. are harmonic.
Therefore by \ref{prod2}.d, so is $\T_{\bar n}$ provided that
$n_i\equiv 0 \!\!\mod    5$ for at least $2$ values of $i$, or
$n_i\equiv 0 \!\!\mod    7$ for at least $3$ values of $i$, etc.
However, $C_5$, $\T_{(7,7)}$ and $\T_{(11,11)}$ are not harmonic,
see \ref{dpo}.a, \ref{mers} and Appendix 1 below. \eexa

\subsection{Partners} 
\bsit \label{cub1} Let $E$ be the affine plane cubic with equation 
\be\label{cueq} (1 +x+y)(1 + xy)=
1\,,\ee and set $E^*=E\setminus \{ (0,0)\}$.
By virtue of \ref{q-1}.b the  toric 2-lattice
$\T_{m,n}$ is harmonic if and only if the curve
$E^*(\bar k)$ possesses a point $(x,y)$ with 
$ \ord\, x \mid m$
and $\ord\,y \mid n$.

We are interested in the infinite table
$\cE$ composed of all pairs $(m,n)\in\N^2$ such that the lattice
$\T_{m,n}$ is harmonic, or equivalently, such that
$x^m=y^n=1$ for some point $(x,y)\in E^*(\bar k)$. We consider 
also the subtable
$$\cE_0=\{(\ord\,x, \ord\,y)\,:\,  (x,y)\in E^*(\bar k)\}\subseteq \cE\,.$$
Thus $\cE_0$ is the set of all bi-torsions of points on $E^*$, or
in other words, the set of all minimal bi-periods of double
periodic binary harmonic functions on $\Z^2$. Notice that $\cE_0$
contains a set of primitive generators of $\cE$ viewed as a module
over the multiplicative semigroup $\N^2$ (cf.
\ref{prod2}.d). We call $m$ and $n$ {\it
partners} if $(m,n)\in\cE_0$. For instance $(47,\,178481)$ 
is a pair of partners found by Zagier.

The following lemma is a reformulation of Theorem 5.2 in \cite{HMP}; 
the latter
also 
covers the case of square grids over $\F_3$. \esit

\bprop\label{infpr} The table of partners $\cE_0$ is infinite. 
\eprop

\bproof If $\cE_0$ were finite there would exist a prime $p$ such
that $$p>M=\max\{m\,:\, (m,n)\in \cE_0\,\,\mbox{for
some}\,\,n\in\N\}\ge 5\,.$$ We have $d\equiv 1 \!\!\mod    p$
$\forall d\mid (2^p-1)$. Indeed, for every prime divisor $l$ of
$2^p-1$,
$$2^p\equiv 1 \!\!\mod    l 
\,\,\,\Longrightarrow \,\,\, \ord_l\, 2=p 
\,\,\,\Longrightarrow \,\,\, l\equiv 1 \!\!\mod p\,.$$ 
It follows that $d\equiv 1 \!\!\mod    p$  and so, $d>p>M$ if
$d>1$.

However the toric lattice $\T_{q-1,q-1}$ being harmonic by
\ref{dppl}, we must have $(q-1,q-1)=(kd_1,ld_2)$ for some
$d_1,\,d_2,\,k,\,l\in\N$ such that $(d_1,d_2)\in\cE_0$ and
$d_1>1$. Since $d_1>M$ this yields a contradiction.
\eproof

Somewhat more precise information can be deduced by using the Hasse-Weil 
formula.

\subsection{Hasse-Weil formula}  The cubic curve $E$ as in \ref{cub1} has 
3 points at
infinity: $(1:0:0), \,(0:1:0)$ and $(1:1:0)$.
Hence the projective closure $\bar E$
of $E$ is a smooth elliptic curve. For $q=2^r$ we let $\bar
s_r$ ($s_r=\bar s_r-4$, respectively) be the number of points on 
$\bar E(\F_q)$ ($E^*(\F_q)$, respectively). 

\blem\label{HW} We have
$s_r=q\left(1-(\alpha_+^r+\alpha_-^r)\right)-3$, where
$\alpha_\pm=(-1\pm\sqrt{-7})/4$ are the complex roots of the
polynomial $2t^2+t+1$. Moreover, the Hasse inequalities hold:
$$({\sqrt{q}}-1)^2\le \bar s_r \le
({\sqrt{q}}+1)^2\,.$$\elem

\bproof Since $\bar s_1=4$, the Hasse-Weil formula \cite[Ch. V,
\S 1, Exercise 7]{Ko} gives in our case:
$$\sum_{r=1}^{\infty} \frac{\bar s_r}{r} t^r= \log\, \zeta_{\bar E} (t)
=\log \frac{1+t+2t^2}{(1-t)(1-2t)}=\sum_{r=1}^{\infty}
\frac{t^r}{r} \left(1+2^r(1-(\alpha_+^r+{\alpha}_-^r)) \right)\,.
$$
Now the assertions follow easily.\eproof

From \ref{HW} and \cite[\S 5, Remark]{HMP}
we deduce the following results.

\bcor\label{mers} \begin{enumerate}\item[(a)]
$\forall q=2^r\ge 16$, $E^*(\F_q)\neq\emptyset$.
\item[(b)] All Mersenne primes $q-1=2^p-1$ with
$p>3$, and all Fermat primes $q+1=2^{2^l}+1$ with $l\ge 1$,
are self-partners i.e., $(q\pm 1,\, q\pm 1)\in\cE_0$. 
Whereas for $p=2,3$
one has $(3,3),\,(7,7)\not\in \cE_0$.
\end{enumerate}\ecor

\bproof (a) follows by virtue of \ref{HW}.

(b) According to \ref{dppl}, the toric lattice $\T_{q\pm 1,q\pm
1}$ is harmonic $\forall q=2^r$, except for $\T_{1,1}$ and
$\T_{7,7}$. Thus for all those $m=n=q\pm 1$, (\ref{ssy}) has a
solution $(\xi, \eta)$. If $q\pm 1$ is prime and $\xi,\eta\neq 1$
then $\ord\,\xi=\ord\, \eta=q\pm 1$ that is, $(q\pm 1,q\pm 1)\in
\cE_0$. This proves the first assertion.

Since $x= 1$ or $y= 1$ for every point $(x,y)\in E^*(4)$ then
$(\ord\,x, \ord\,y)=(1,3)$ or $(3,1)$  and so $(3,3)\not\in \cE_0$. Neither 
$(7,7)\in \cE_0$ since $s_3=0$. \eproof

So far only $5$ Fermat primes and at most $43$ Mersenne primes
were found, see e.g., \cite[7.3]{LLMP}, \cite{Me}, \cite{Wa}.

The computer findings in Appendix 1
suggest the following conjecture, cf. \ref{dppl}:

\bconj\label{conje} $\forall q=2^r$ ($r\ge 6$), $q-1$ and $q+1$ are
partners and auto-partners that is, $(q\pm 1, q\pm 1)\in \cE_0$ and
$(q+1,q-1)\in\cE_0$.\econj

\noindent The latter does not hold for $r=5$. Indeed $(31,33)\in \cE\setminus\cE_0$,
see  \ref{dppl} and Appendix 1. 

\bexas\label{57n} 1. $(5,n)\in \cE$ if and only if $n\equiv 0
\!\!\mod   3$ or $n\equiv 0 \!\!\mod   5$. In particular 
$(5,5)\in \cE_0$ is a self-partner, and there is no further partner of $5$.

Indeed, as $(1,3)\in \cE$ then $(k,3l)\in \cE$ for every $k,l\ge
1$, in particular for $k=5$. Further, $5$ is a primitive
self-partner since for any primitive $5$-th root of unity
$\zeta\in \mu_5$, the pair $(x,y)=(\zeta,\zeta^2)$ satisfies
(\ref{ME}) with $s=2$, $m=n=5$. Consequently by virtue of
\ref{prod2}.a, $(5k,5l)\in \cE$ $\forall k,l\ge 1$  and so
$(5,5l)\in \cE$ $\forall l\ge 1$.

Conversely, if $(x,y)\in E^*$ and $(\ord\,x, \ord\,y)=(5,n)$ then
$x\in \F_{16}\setminus \F_4$ satisfies $x^5= 1$ and
$y=x^2,\,x^{-2}$ satisfy
$$y^2+(z+1)y+ 1= 0,\quad\mbox{ where} \quad z=x+x^{-1}\,.$$ Thus
$n= \ord\,y=5$.

2. $(7,n)\in \cE$ if and only if $n\equiv 0 \!\!\mod   3$, and $9$
is the only partner of $7$.

Indeed, $(7,3k)\in \cE$ $\forall k\ge 1$ because $(1,3)\in \cE$. If
$(x,y)\in E^*$ and $(\ord\,x, \ord\,y)=(7,n)$ then
$x\in\F_8\setminus\F_2$. We have 
$(7,n)\in\cE_0\quad\Longrightarrow\quad
f_0(n)=f_0(7)=3\quad\Longrightarrow\quad n\mid (2^3\pm
1)\quad\Longrightarrow\quad n\in\{7,9\}$. But $n\neq 7$ as $s_3=0$
and so, $E^*(\F_{2^7})=\emptyset$. Hence $n=9$.

Since $(1,3),\,(7,9)\in \cE_0$, the latter set properly
contains the set of all primitive generators of $\cE$ over $\N^2$.
\eexas

\subsection{Partnership graph}  We observe that:

\medskip

\begin{enumerate}
\item[$\bullet$]
For every $(m,n)\in \cE_0$, both $m$ and $n$ are odd (cf.
\ref{prod2}.c).
\item[$\bullet$]
Every odd $n\in\N$ has a partner, and the number of these
partners is finite.
\item[$\bullet$]
$(m,n)\in \cE_0 \iff (n,m)\in \cE_0$.
\end{enumerate}

\medskip

\noindent Thus the partnership defines an equivalence
relation on $\N_{\rm odd}$. Answering a question of the 
author, Zagier proposed the following \ref{pargra}, \ref{MT} and
\ref{ki} below. Our proof of \ref{MT} based on \ref{ML} is somewhat
different from the original one.

We let below $\div (n)$ ($\div^* (n)$, respectively) be the set
of all (proper) divisors of $n\in\N$. For $q=2^r$ we write for
short $\div (q\pm 1)$ meaning $\div (q-1)\cup \div (q+1)$.

\bdefi\label{pargra} We let ${\mathcal P}^{(1)}$ be the infinite
graph with loops such that $\ver ({\mathcal P}^{(1)})=\NO$ and
$[m,n]\in {\edg} ({\mathcal P}^{(1)})\iff (m,n)\in \cE_0$. We
call ${\mathcal P}^{(1)}$ the {\it partnership graph}.
\edefi

\bthm\label{MT} All connected component of ${\mathcal P}^{(1)}$
are finite. \ethm

\bproof We let ${\mathcal V}_r$ be the subgraph of ${\mathcal
P}^{(1)}$ with vertices in the finite set
$$V_r=\{n\in\NO\,:\, n\mid (2^r\pm 1)\}\,.$$
Given $n\in\NO$, we let
${\mathcal P}^{(1)}(n)$ be the connected component of ${\mathcal
P}^{(1)}$ which contains the vertex $n$. We claim that the
function $f_0(n)$ is constant on each connected component of
${\mathcal P}^{(1)}$. In particular ${\mathcal
P}^{(1)}(n)\subseteq {\mathcal V}_{f_0(n)}$. The level sets $V_r$
of $f_0$ being finite, this proves the theorem.

To show the claim we note that, due to \ref{q-1}.a, $[m,n]\in\edg
({\mathcal P}^{(1)})$ if and only if
$\xi+\xi^{-1}=1+\eta+\eta^{-1}$ for some primitive roots
$\xi\in\mu_m$ and $\eta\in\mu_n$. According to \ref{fod}.a,
$$f_0(m)=\deg (\xi+\xi^{-1})=\deg (\eta+\eta^{-1})=f_0 (n)\,$$
 and so, the claim follows. \eproof

\bnota\label{ki} We denote by $S (m,n)$ the set of all 
solutions $(\xi,\eta)$ of (\ref{cueq}) of type $(m,n)$ that is,
with $\xi\in \mu_m$ (respectively, $\eta\in \mu_n$) being a
primitive $m$-th (respectively, $n$-th) root of unity. We label
the edges $[m,n]\in\edg ({\mathcal P}^{(1)})$ with
$s(m,n)=\frac{1}{2}\card (S_{m,n})\in\N$, with one exception:
instead of the edge $[1,3]$ we introduce two directed edges,
$[1\to 3]$ labeled by $1$ and $[3\to 1]$ labeled by $2$. Clearly,
$$s_r=2\sum_{m,n\in\div (q-1)} s(m,n),\qquad \forall r\in\N\,.$$
Moreover this labeling possesses the following properties. \enota

\bprop\label{euler} $\forall n\in\N_{\rm odd}$ and $\forall
q=2^r,\,r\ge 3$, 
\begin{enumerate} \item[(a)] $\sum_{m\in\N_{\rm odd}}
s(m,n)=\varphi(n)$ \footnote{Hereafter 
$\varphi$ stands for the Euler function.}.
\item[(b)] $\sum_{d\in\div(n),\,m\in\N_{\rm odd}} s(d,m)=n$.
\item[(c)] $\sum_{n\in V_r}\varphi (n)=2\sum_{m,n\in V_r,\;m\neq n}
s(m,n)+ \sum_{n\in V_r} s(n,n)$.
\item[(d)] $2\sum_{d, d'\in\div (q\pm 1),\; d\neq d'}
s(d,d')+ \sum_{d\in\div (q\pm 1)} s(d,d)=2q\,.$
\item[(e)]
$s(q-1,q-1)+s(q+1,q+1)+2s(q-1,q+1)\ge 2\left(\varphi (q-1)
+\varphi (q+1)-q\right)$.
\end{enumerate}\eprop

\bproof (a) holds because for every $(\zeta,\eta)\in S (m,n)$, the
pairs $(\eta, \eta^{-1})$ and $(\zeta, \zeta^{-1})$ uniquely
correspond to each other. Since $ \sum_{d\mid n}\varphi (d)=n$,
(b) follows from (a). Summing up (a) over the edges of ${\mathcal
 V}_r$ yields (c). It is easily seen that $d\in\div (q\pm
1)\quad\iff\quad f_0(d)\mid r \quad\iff\quad d\in V_s$ for some
$s\in\div (r)$. Hence $$\sum_{s\mid r} \sum_{n\in V_s} \varphi
(n)=\sum_{n\in \div (q\pm 1)} \varphi (n)=\sum_{n\mid (q-1)}
\varphi (n)+\sum_{n'\mid (q+1)} \varphi (n')=2q\,.$$ Thus the
summation of (c) over the set $\div (r)$ yields (d).

(e) By virtue of \ref{euler}.a, $s (d,q-1)+s (d,q+1)\le \varphi
(d)$. Moreover
$$\varphi (q-1)+\varphi (q+1)=s(q-1,q-1)+s(q+1,q+1)+2s(q-1,q+1)$$
$$+\sum_{d\in\div^* (q\pm
1)} \left(s(d,q-1)+s(d,q+1)\right)\,.$$ Hence
$$s(q-1,q-1)+s(q+1,q+1)+2s(q-1,q+1)\ge
\left(\varphi (q-1)-\sum_{d\in\div^* (q-1)} \varphi (d) \right)$$
$$ + \left(\varphi (q+1)-\sum_{d'\in\div^* (q+1)} \varphi
(d')\right)=2\left(\varphi (q-1) +\varphi (q+1)-q\right)\,.
$$\eproof

\brems\label{gn} 1. If $(\zeta,\eta)$ is a solution of
(\ref{cueq}) of type $(m,n)$ then $(\zeta^2,\eta^2)$ is as well
such a solution. Thus the Galois group ${\rm Gal} (\F(\zeta))$
acts freely on $S(m,n)$  and so, its order $f(n)$ divides
$s(m,n)$. Letting $k(m,n)=s(m,n)/f(n)$, from \ref{euler}.a we
obtain the equality
$$\sum_{m} k(m,n)=g(n)\,.$$ Indeed, there are $\varphi (n) =
f(n)g(n)$ primitive $n$-th roots of unity.

2. For $q=2^r$ the
inequality 
$\varphi (q-1) +\varphi (q+1)>q$ does not hold in general,
although it holds at least for all $r\le 150$
(a MAPLE checking)\footnote{Computations with Pari/GP
done by Gottfried Barthel (a letter to the
author) confirm the inequality in the range $r\le 275$.}. 
Moreover, according to A.\ Schinzel\footnote{A letter to the
author.}, 
$\varphi  (2^r -1) +
\varphi  (2^r +1)$ can be $<< 2^r/(\log\log r_t)^{7/24}$ 
for infinitely many $r$, although
$\varphi (2^r - 1) + \varphi (2^r + 1)\ge 2^r/\log\log r$ $\forall r\in \N$. 
This can be seen as follows.
Let
$r_t$ be the least common multiple of all numbers $(p-1)/2$,where $p$ 
runs through primes $\equiv 3 \mod 4$ less than $t$. By Euler's Theorem and 
the quadratic reciprocity law, $p$ divides $2^{(p-1)/2} - (-1)^{(p+1)/4}$,
hence $2^{r_t} -1$ is
divisible by the product of all primes $\equiv 7 \mod 8$ less than $t$,
while $2^{r_t} +1$
is divisible by the product of all primes $\equiv 3 \mod 8$ less than $t$.
Using the Mertens formula for primes in arithmetic progressions, we obtain that
$\varphi  (2^{r_t} \pm 1)<< 1/\log t << 1/(\log\log r_t)^{1/4}$. A slight modification 
of the argument increases the exponent $1/4$ to $7/24$, which gives the claim.
But a gap between the lower and the upper bound remains.
Inded from a theorem of Erd\"os \cite{Er}, 
and from the inequality between phi and sigma functions 
\cite[Theorem 329]{HW} it follows that
$\varphi ( 2^r -1) + \varphi (2^r +1) >> 2^r/\log\log r$.


3. Similarly to ${\mathcal P}^{(1)}$, 
one might consider an infinite hypergraph 
${\mathcal P}=\cup_{s\ge 1} {\mathcal
P}^{(s)}$ with set of vertices
$\N_{\rm odd}$ such that $(n_0,\ldots,n_s)$ is an $s$-simplex of
${\mathcal P}^{(s)}$ if and only if $(n_0,\ldots,n_s)$ is the multi-order of 
a point on the affine hypersurface as in (\ref{ME}) (with
$s$ replaced by $s+1$). Evidently, every $s$-tuple
$(n_1,\ldots,n_{s})\in \N_{\rm odd}^{s}$ is a face of an $s$-simplex
in ${\mathcal P}^{(s)}$. \erems

\section{Appendix 1: Connected components of the partnership graph}
The first 13 connected components of the partnership graph
${\mathcal P}^{(1)}$ are shown below. They were 
found by Zagier with PARI. The labeling of the edges is according to \ref{ki}. 
We recall (see the proof of \ref{MT}) that the
value of $f_0(n)$ equals $r$ for every vertex $n$ of ${\mathcal
V}_r$, and this determines ${\mathcal
V}_r$. The value of $f(n)$ equals $r$ if $n$ is not
underlined, and $2r$ otherwise. The edges $[m,n]$ 
(the loops $[n,n]$, respectively)
correspond to harmonic toric lattices $\T_{m,n}$ ($\T_{n,n}$, respectively) with 
$(m,n)\in \cE_0$. 
\newpage
$\,$
\vskip 1in

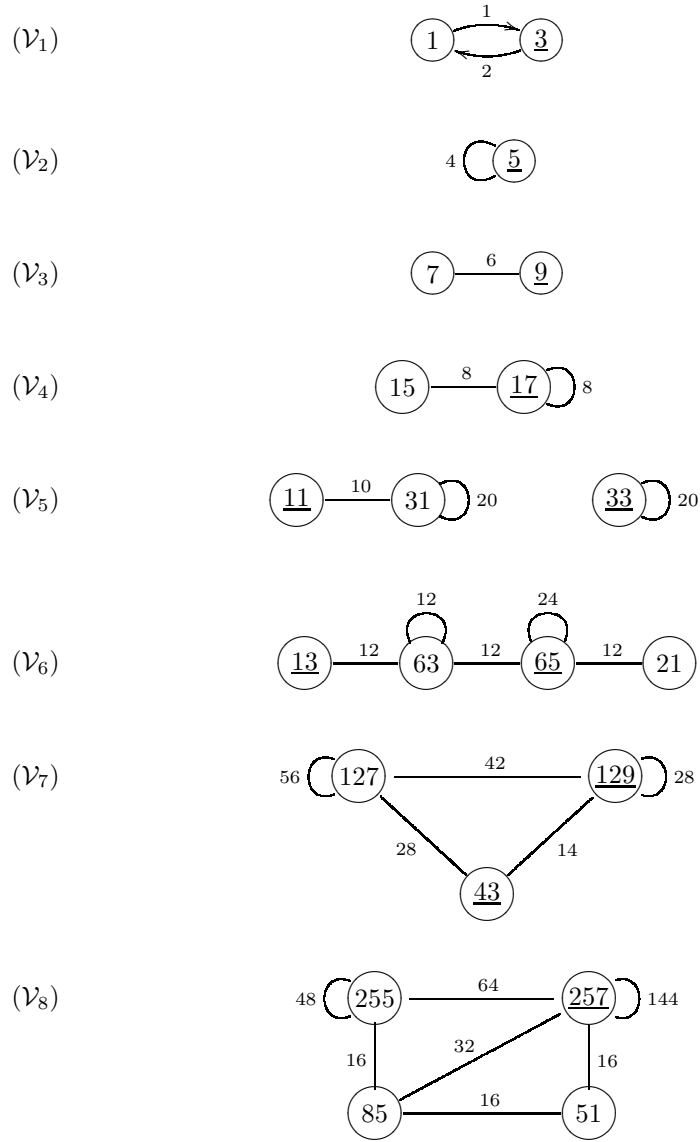
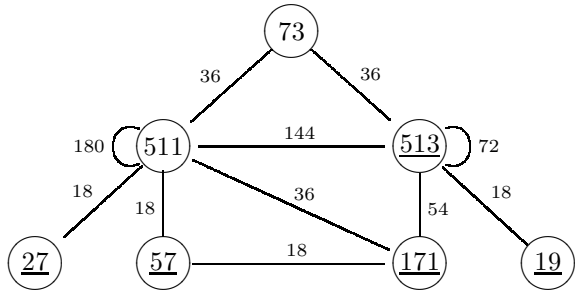
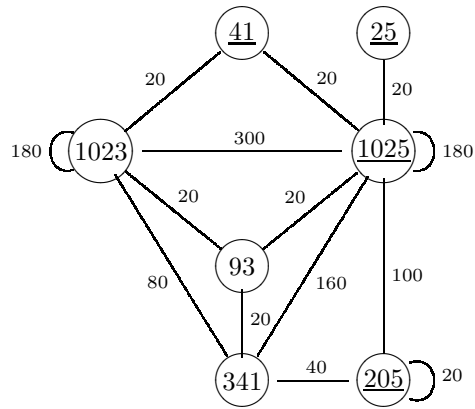
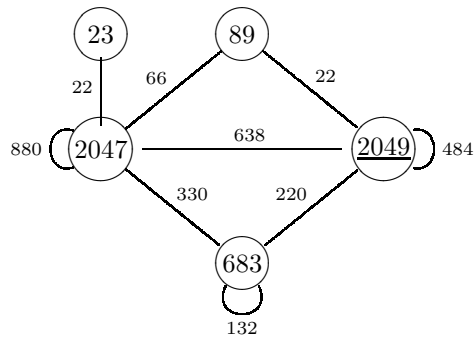
\begin{center}
\includegraphics{hawo1eps.epsi}
\end{center}

\newpage 
$\,$
\vskip 1in

\begin{center}
\includegraphics{hawo2eps.epsi}
\end{center}

\newpage 
$\,$
\vskip 1in

\begin{center}
\includegraphics{hawo3eps.epsi}
\end{center}


\noindent We observe that the graph ${\mathcal V}_{12}$ is not planar, 
in contrast to ${\mathcal V}_{r}$ with $r\le 11$. 

\noindent These computations suggest the following

\noindent \bconj\label{zconj} ${\mathcal V}_r$ is connected $\forall r\neq
5$. In other words, the connected components of ${\mathcal
P}^{(1)}$ are ${\mathcal V}_r$ for $r\neq 5$ and the two components of 
${\mathcal V}_5$. \econj

\section{Appendix 2: Chebyshev-Dickson and Fibonacci polynomials}
\label{CDF}
\subsection{Chebyshev-Dickson and Fibonacci polynomials}
These polynomials $T_n,\,E_n$ and $F_n$ provide an important tool
for analysis of harmonicity. Indeed, as we have seen in
\ref{charpo}, $T_n$, respectively, $E_n$ is the characteristic
polynomial of the laplacian $\Delta^-_{C_n}$, respectively,
$\Delta^-_{P_n}$, where $C_n$ stands for the circular graph with
$n$ vertices and $P_n$ denotes the path of length $n$. 
We give an account of
some of their properties in \ref{dpo}-\ref{dpo11} below according
to \cite[Ch. 2]{LMT}, \cite{Su3}, \cite{BR}, \cite{GKW}, see also
references therein.

\bdefi\label{dpo} Consider the linear recurrence \be\label{recur}
p_{n+1}=xp_n+p_{n-1},\qquad\mbox{where}\quad p_i\in
k[x]\quad\forall i\ge 0\,.\ee Thus
$$\left(
\begin{array}{c} p_n \\ p_{n+1}
\end{array}\right)
=\left(
\begin{array}{cc}
{ 0} & { 1} \\ { 1} & x
\end{array}
\right)^n \left(
\begin{array}{c} {p_0} \\ {p_1}
\end{array}
\right)\,.
$$
The {\it Chebyshev-Dickson polynomials} of the first,
respectively, second kind $T_n,\,E_n\in k[x]$ and the {\it
Fibonacci polynomials} $F_n\in k[x]$ are defined via
(\ref{recur}) by the initial conditions
$$\left(
\begin{array}{c} T_0 \\ T_{1}
\end{array}\right)
=\left(
\begin{array}{c} { 0} \\ x
\end{array}
\right),\qquad \left(
\begin{array}{c}
E_0 \\ E_{1}
\end{array}
\right) =\left(
\begin{array}{c}
1 \\ x
\end{array}
\right),\qquad \left(
\begin{array}{c}
F_0 \\ F_{1}
\end{array}
\right) =\left(
\begin{array}{c}
0 \\ 1
\end{array}
\right), \quad\mbox{respectively}.$$ Thus $\deg T_n=\deg E_n=\deg
F_{n+1}=n$, the polynomials $T_{2n},\,E_{2n},\,F_{2n+1}$ are even
and $T_{2n+1},\,E_{2n+1},\,F_{2n}$ are odd. They are related via
\be\label{dpe} T_n=xF_n=xE_{n-1},\qquad\mbox{ where}\quad
E_{-1}=0\,.\ee So any property of one of the sequences
$(T_n),\,(E_n),\,(F_n)$ is enjoyed by the other two up to evident
changes. Notice that $F_n(1)$ is the $n$-th Fibonacci number
modulo $2$. \edefi

The
following identities hold, see e.g. \cite{Ri, LMT}, \cite[\S 4]{BR},
\cite{WP}.

\bprop\label{dpo1} $\forall m,n\in\N$, $\forall q=2^r$, $r\ge 1$, we have
\begin{enumerate}\item[(a)] $F(z)=z
(z^2+xz+1)^{-1}$ is the generating function of the sequence
$(F_n)$.
\item[(b)] 
$$E_n(x)=\sum_{i=0}^{[n/2]} \binom{n-i}{i} x^{n-2i} \!\!\! \mod  2 =
\sum_{j=0,\ldots,n,\,\,j\equiv n\! \!\!\!\mod  2} \binom{n+j}{n-j}
x^{j} \!\!\!\mod  2 \equiv U_n\left(\frac{x}{2}\right)\!\!\!\mod
2\,,$$ where $U_n(\cos x)=\frac{\sin nx}{\sin x}$ stands for the
Chebyshev polynomial of the second kind over $\R$.
\footnote{Attention: our enumeration of classical polynomials
does not coincide with those used in MAPLE. It is so chosen in order to write
the identities in a more elegant way.}
\item[(c)] $F_{q-1}+F_{q+1}=xF_q=x^q$. 
Furthermore,
$$F_{q+1}(x)=x^q+F_{q-1}(x)=x^q\left(1+\sum_{i=0}^{r-1}
x^{-2^i}\right)^2\,.$$
\item[(d)] 
$\forall z\in\F_q^\times$,
$$F_{q-1}(z)=z\Tr_{\F_q}^2 (z^{-1}),\quad F_{q+1}(z)=z\left(1+\Tr_{\F_q}
(z^{-1})\right)^2\quad\mbox{and}\quad F_{q-1}(z)+F_{q+1}(z)=z\,.$$
\item[(e)] $E_{m+n}=E_mE_n+E_{m-1}E_{n-1}$.
\item[(f)]
$E_{2n}=xE_nE_{n-1}+1=E_n^2+E_{n-1}^2$, $E_{2n+1}=xE_n^2$
and $\sum_{i=1}^n E_{2i}=E_n^2$.
\item[(g)] $\forall n\ge t\ge 0$,
$T_{n+t}+T_{n-t}=T_nT_t$.
In particular for $n\equiv t \!\!\mod  2$,
$T_n+T_t=T_{\frac{n+t}{2}}T_{\frac{n-t}{2}}$.
\item[(h)] $T_m\circ T_n=T_n\circ T_m=T_{mn}$.
\item[(i)] $T_{qn}(x)=T_n(x^q)=T^q_n(x)$.
\end{enumerate}\eprop

\bproof If $\G=P_n$  is a linear graph with $n$ vertices and
$e=n-1$ edges then $\left(\frac{e}{i}\right)_\G=\binom{n-i}{i}$. So
(b) follows by virtue of \ref{grdet}.a and \ref{charpo},
whereas \ref{grdet}.c yields (e). In turn (e) implies (f), (g).
Now (c) follows by recursion and implies (d). 
The assertions (a), (h) and (i) can be deduced, by virtue of (b), 
from the analogous
identities for the usual Chebyshev polynomials over $\R$, see
e.g., \cite{LMT, Ri}. \eproof

\bsit\label{NO}
As before, $\ord\,\xi\in\NO$ denotes the
multiplicative order of an element $\xi\in\bar k^\times$. 
In the next proposition we indicate certain divisibility
properties and factorization of the Chebyshev-Dickson and
Fibonacci polynomials according to \cite[Ch. 2]{LMT}, \cite{Su3},
\cite{GKW}, \cite[\S 4]{BR} and \cite{WP}. \esit

\bprop\label{dpo11} \begin{enumerate}
\item[(a)] $\forall n\ge 0$, $\forall \xi\in\bar k^\times$,
$T_n(\xi+\xi^{-1})=\xi^n+\xi^{-n}$.
\item[(b)] $\forall\xi\in\bar k^\times$, $\ord\,
\xi=\min\{n>0\,:\,
T_n(\xi+\xi^{-1})= 0\}$.
\item[(c)] $\forall n\in\NO$ and for any
 primitive $n$-th root of unity $\zeta\in\mu_n$, we have
$$T_n (x)=x\prod_{i=1}^{(n-1)/2} (x+\zeta^i+\zeta^{-i})^2\,.$$
Consequently, $z=\xi+\xi^{-1}$ runs over the roots of $T_n$ when
$\xi$ runs over $\mu_n$. \footnote{By making use of \ref{dpo1}.h,
one can deduce in the same way the
roots of $T_n$ for any even $n$.}
\item[(d)] $E_{n-1}(0)=0\quad\iff\quad n\equiv 0  \!\!\!\mod  2$,
$E_{n-1}(1) =0\quad\iff \quad n\equiv 0  \!\!\!\mod  3$ and\\
$(x^2+x+1)\mid E_{n-1}\quad\iff\quad n\equiv 0 \!\!\mod  5$.
\item[(e)] $\forall m,n\in\N$,
$\gcd (T_m,T_n)=T_{\gcd (m,n)}$ and $\gcd
(E_{m-1},E_{n-1})=E_{\gcd (m,n)-1}$.
\item[(f)]
$T_d\mid T_n \quad\iff\quad E_{d-1}\mid E_{n-1}\quad\iff\quad
d\mid n$.
\end{enumerate}\eprop

\subsection{Irreducible factors of Fibonacci polynomials}

\bsit\label{fz} Every $z\in\bar k$ can be written in a unique way
as $z=\zeta+\zeta^{-1}$, where $\zeta$ and $\zeta^{-1}$ are the
roots of $f_z(x)=x^2+zx+1\in\bar k[x]$.  By virtue of
\ref{dpo11}.c, every irreducible polynomial $\tau\in k[x]$ divides
one of the $F_n$ \cite[3.1]{Su3}. Namely, if $\tau\neq x$ and
$\tau(\zeta+\zeta^{-1})=0$ then $\tau\mid F_n$ with $n=\ord\,
\zeta$. \esit

\brem\label{write}  An element 
 $z\in\F_q^\times$ ($q=2^r$) can be written as $z=(u^2+u)^{-1}$ for
some $u\in\F_q\setminus\F_2$ if and only if it can be written as
$z=\xi+\xi^{-1}$ for some $\xi\in \F_q^\times$, 
see \ref{dpo1}.d and \ref{dpo11}.c.  \erem 

These observations lead to the following definition \cite{GKW,
Su3}.

\bdefi\label{FI} Letting $\Irr[x]$ be the set of all 
irreducible polynomials in $k[x]$, we remind that for $\tau\in
\Irr[x]$, $\ord\,\tau=\ord\, z\in\NO$, whenever $z$ is a root of
$\tau$. For $\tau\neq x$ we define its {\it Fibonacci order}
\footnote{$\ford\,\tau$ is called {\it Fibonacci index} of
$\tau$ in \cite{GKW}, and {\it depth} of $\tau$ in
\cite{Su3}.}
$$ \ford\, \tau=\ford\, z=\min\{n>0\,:\, \tau\mid F_n\}=\ord\,
\zeta\in\NO,\quad\mbox{where}\quad z=\zeta+\zeta^{-1}\,.$$\edefi

\bsit\label{cyclot} For $n\in\NO$,
$x^n-1=\prod_{d\mid n} \Phi_d$ is a product of cyclotomic
polynomials \be\label{cyclo}
\Phi_d(x)=\prod_{\tau\in\Irr[x],\,\ord\, \tau=d} \tau
=\prod_{1\le i\le d-1,\,\gcd (i, d)=1} (x-\zeta^i)\,,\ee where
$\zeta\in\mu_d$ is a primitive $d$-th root of unity. Hence $\deg
\Phi_d=\varphi (n)$, and $\Phi_d$ is a product of $g(d)=\varphi
(d)/f(d)$ distinct irreducible factors of the same degree $f(d)$
\cite[2.47]{LN}.
By virtue of (\ref{cyclo}), $\Phi_d$ is auto-reciprocal, that is
$\Phi_d^*=\Phi_d$, where
$$* : p(x)\longmapsto x^{\deg p} p(x^{-1})$$ is an involutive
automorphism of the multiplicative semigroup
$$\Pi[x]=\{p\in k[x]\,:\, p(1)=1\}\,.$$
It occurs that either all  irreducible factors of $\Phi_d$ are
auto-reciprocal or none of them is, depending on $d$. More
precisely, the following happens. \esit

\blem\label{ML} Let $\zeta\in\mu_d$ be a primitive root of unity
of odd order $d$. We denote by 
$\tau (x)=\prod_{j=0}^{r-1} (x-\zeta^{2^j})\in k[x]$ 
its minimal polynomial of
degree $r=f(d)$. 
The following conditions are equivalent:
\begin{enumerate}
\item[(i)] $\tau$ is auto-reciprocal, or palindrome.
\item[(ii)] $2^{f_0(d)}\equiv -1\mod d$.
\item[(iii)] $\deg\zeta=2\deg (\zeta+\zeta^{-1})$.
\footnote{Or, equivalently, $[\F_2(\zeta) :
\F_2(\zeta+\zeta^{-1})]=2$.}
\item[(iv)] ${\Tr}_{k(\zeta+\zeta^{-1})}
(\zeta+\zeta^{-1})^{-1} =1$.
\end{enumerate}
\elem

\bproof $\tau$ being irreducible of degree $\ge 2$, we have $\tau
(1)=1$. But $1$ is the only fixed point of the involution
$z\longmapsto z^{-1}$ on $\bar k^\times$. In case that
$\tau=\tau^*$ this involution acts on the roots of $\tau$, hence
$f(d)=\deg\tau$ is even.

The roots of $\tau$ being the conjugates of $\zeta$ in $\bar  k$,
$\tau$ is auto-reciprocal if and only if $\zeta$ and $\zeta^{-1}$
are conjugated.  By virtue of \ref{subord} this yields the equivalence
(i)$\Longleftrightarrow$(ii). The condition (iii) holds if and
only if the polynomial $f_z$ as in \ref{fz} above is irreducible,
where $z=\zeta+\zeta^{-1}$. Thus the equivalence
(iii)$\Longleftrightarrow$(iv) follows, see e.g.
\cite[8.13]{McE}. To show the remaining equivalence
(i)$\Longleftrightarrow$(iii) we consider the Laurent polynomial
$g(x)=\tau(x)\tau(x^{-1})\in k[x,x^{-1}]$. It is {\it
auto-reciprocal} that is, $g(x)=g(x^{-1})$ or, equivalently,
$g(x)=h(x+x^{-1})$ for some $h\in k[z]$ of degree $r$. As
$h(\zeta+\zeta^{-1})=g(\zeta)=0$, (iii) holds if and only if $h$
is reducible.

Supposing (i) we have $r=2s$, where $s\in \N$, and
$\tau(x^{-1})=x^{-r}\tau(x)$. Thus
$h(x+x^{-1})=\left(x^{-s}\tau(x)\right)^2=:\tilde g^2 (x)$. The
Laurent polynomial $\tilde g$ being auto-reciprocal, it follows
that $\tilde g(x)=\tilde h (x+x^{-1})$, where $\tilde h\in k[x]$,
$\deg\tilde h=s$ and $h={\tilde h}^2$. Clearly, $\tilde h$ is
the minimal polynomial of $\zeta+\zeta^{-1}$. This yields (iii).

Conversely, let (iii) holds i.e., $h=h_1h_2$ is reducible, where
$h_i\in k[x]$, $\deg h_i=r_i\ge 1$, $i=1,2$ and $r_1+r_2=r$.
Letting $g_i (x)=x^{r_i}h_i(x+x^{-1})\in k[x]$ we have $\deg g_i
=2r_i$, $i=1,2$. Furthermore
$$\tau(x)\tau^*(x)=x^r \tau(x)\tau(x^{-1})=x^r h(x+x^{-1})=
g_1 (x)g_2 (x)\,.$$ Since $\tau, \,\tau^*\in\Irr[x]$, up to
interchanging $g_1$ and $g_2$ we obtain $g_1=\tau,\,\,g_2=\tau^*$.
Hence $r=2r_1=2r_2$ is even and
$$\tau(x)=g_1 (x)
= x^{r/2} h_1 (x+x^{-1})= x^r \tau(x^{-1})=\tau^*(x)\,.$$ Thus
$\tau=\tau^*$, so (i) holds. \eproof


\bcor\label{fod} \begin{enumerate}\item[(a)] For every primitive
$d$-th root of unity $\zeta\in\mu_d$ we have $\deg\zeta=f(d)$
and $\deg (\zeta+ \zeta^{-1})=f_0(d)$.
\item[(b)] Let $\tau(x)=\sum_{i=0}^r \varepsilon_i x^i$
be the minimal polynomial of $\zeta$, and let  $\tau (x)\tau
(x^{-1})=1+\sum_{j=1}^r \delta_{j} (x^j+x^{-j})$. Then the
minimal polynomial $\eta(x)$ of $\zeta+\zeta^{-1}$ is
$$\eta(x)=\begin{cases}
1+\sum_{j=1}^s \varepsilon_{s-j} T_j(x) &
\\
\\1+\sum_{j=1}^r \delta_{j} T_j(x) &\end{cases}
\qquad\mbox{with}\qquad \deg\eta=\begin{cases} s=r/2 \quad
\mbox{if}\quad \tau=\tau^*,\\ r \quad \mbox{otherwise}
\,.\end{cases}$$
\end{enumerate}\ecor

Following \cite{GKW}, \cite[3.2]{Su3}, \cite{SB} we list below
some important features of irreducible factors of the
Fibonacci polynomials.

\bprop\label{irfib} For every $\tau\in\Irr^*[x]$ with $\ford\,
\tau=d$, the following hold.
\begin{enumerate} \item[(a)] $\deg\tau=f_0(d)$ i.e.,
 the splitting field of $\tau$ is
$\F_q$ with $q=2^{f_0(d)}$.
\item[(b)] $\tau\mid F_n\quad\iff\quad
d\mid n\quad\iff\quad F_d\mid F_n$.
\item[(c)] $d\mid
(q-1)$ (and so $\tau \mid F_{q-1}$) if and only if the linear term
of $\tau$ vanishes, if and only if the polynomial
$f_z(x)=x^2+zx+1\in \F_q[x]$ splits over $\F_q$, where $z\in
\F_q$ is a root of $\tau$. Otherwise $d\mid (q+1)$ (and so $\tau
\mid F_{q+1}$).
\item[(d)] $\forall n=2k+1\in\NO$,
$$F_n=\prod_{\tau\in\Irr^*[x],\,\ford\, \tau \mid n} \tau^2=
R_k^2\,,$$ where $R_k=F_{k+1}+F_k$, $\deg R_k=k$, $R_k$ is
square-free and contains the monomial $x^{k-1}$. The splitting
field of $F_n$ is $\F_q$, where $q=2^{f_0(n)}$.
\item[(e)] $\forall q=2^r$, $r\ge 1$, the
splitting field of $F_{q\pm 1}$ is $\F_q$. Moreover
$$F_{q-1}F_{q+1}=\prod_{\tau\in\Irr^*[x],\,\deg
\,\tau\mid r} \tau^2=(x^{q-1}+1)^2\,.$$
\item[(f)] For every odd prime $p$ and for every irreducible
factor $\tau$ of $F_p$, $$\deg\tau=\begin{cases}
f(p)\quad &\mbox{if}\quad f(p)\equiv 1 \mod 2 ,\\
f(p)/2\quad & \mbox{otherwise}\,.\end{cases}$$ In particular
$F_p=\tau^2$ if and only if either $f(p)=p-1$ or
$f(p)=\frac{p-1}{2}\equiv 1 \mod 2 $. This cannot happen if
$p\equiv \pm 1 \mod 8,\,p>1$.
\end{enumerate}\eprop

\bproof (a) follows by virtue of \ref{fod}.a, and (b) follows
from \ref{dpo11}.b,f. If $z$ is a root of $\tau$ then by (a),
$z\in\F_q^\times$ and so, by virtue of \ref{dpo1}.d, $F_{q-1}
(z)=0 \quad\iff\quad \Tr_{\F_q} (z^{-1})=0$ and $F_{q+1} (z)=0
\quad\iff\quad \Tr_{\F_q} (z^{-1})=1$. By (b), $d\mid (q-1)$ in the
former case and $d\mid (q+1)$ in the latter one. This yields (c). As
$T_n=xF_n$, (d) and (e) follow easily from \ref{dpo1}.f and
\ref{dpo11}.c. For the proof of (f), see \cite{GKW}. \eproof

\brem\label{tra} We let $\Irr_1[x]=\{\tau\in\Irr[x]\,:\,
\tau=x^{\deg\tau}+x^{\deg\tau-1}+\ldots\}$. Thus
$\tau\in\Irr_1[x]$ if and only if $\tau\in\Irr[x]$ and
$\Tr_{k(z)} (z)=1$ for any root $z=\zeta+\zeta^{-1}$ of $\tau$.
By virtue of (d) above, an odd number of irreducible factors of
$R_k$ belong to $\Irr_1[x]$. Hence for every $n\in\NO$, there
exists $d\mid n$ and a primitive $d$-th root of unity
$\zeta\in\mu_d$ such that
$$\Tr_{k(z)}
(\zeta+\zeta^{-1})=\sum_{i=0}^{f_0(d)-1}
(\zeta^{2^i}+\zeta^{-2^i}) =1\,.$$ \erem

\bsit\label{cp} For any $n\in\NO$, the following analog of the
cyclotomic polynomial $\Phi_n$ (see \ref{cyclo}) were
introduced  in \cite{Su3}:
$$\rho_n=\prod_{\tau\in\Irr^*[x],\,\ford\,\tau=n} \tau^2\,.$$
The following properties of these polynomials were established in \cite{Su3}.
\esit

\bprop\label{spefib} 
\begin{enumerate}\item[(a)] $\deg\rho_n=\varphi (n)$.
Furthermore, $\rho_n$ has $\frac{\varphi (n)}{2f_0(n)}$
irreducible factors, all of the same degree $f_0(n)$, the same
multiplicity $2$ and with the same linear term.
\item[(b)] $\forall q=2^r,\,r\ge 0$,
$$F_n=\prod_{d\mid n} \rho_d\qquad\mbox{and}\qquad
F_{qn}=x^{q-1}F_n^q=x^{q-1}\prod_{d\mid n} \rho_d^q\,.$$
\item[(c)] By the M\"obius inversion formula,
$$\rho_n=\prod_{d\mid n} F_{n/d}^{\mu (d)}\,.$$
\end{enumerate}\eprop

\subsection{$+$-involution} The automorphism group $\Aut
(k[x])$ is isomorphic to $\Z/2\Z$ and consists of the identity and 
the involution 
$$\rho: k[x]\to k[x],\qquad p(x)\longmapsto p^+ (x):=p(x+1)\,.$$
Following \cite{GKW} we call $p^+$ the {\it conjugate} of $p$.
This notion plays an important role in the analysis of harmonicity of
plane grids, see Section 2. The ring of invariants
$k^+[x]:=k[x]^\rho=\ker (\delta)$, where $\delta=\rho+{\rm id}\in\End
(k[x])$, consists of all {\it self-conjugate} polynomials
$p=p^+$. Clearly, $\rho$ preserves degree and
irreducibility.

The following proposition extends Lemma 15 in \cite{GKW}.

\bprop \label{plu}
\begin{enumerate}\item[(a)]
${\rm im} (\delta)=\ker (\delta)=\vect \left(\delta (x^n)\,:\,
n\in\NO\right)$.
\item[(b)] We have $\deg\delta (p)=\deg
p-1$ $\forall p\in k[x]$ with $\deg p\in\NO$, and  $\deg
p\in \NE$ $\forall p\in k^+[x]$.
\item[(c)] $k[x]=k^+[x]\oplus k_{\rm odd} [x]$,
where $k_{\rm odd} [x]\subseteq k[x]$ is the subspace of odd
polynomials.
\item[(d)] $\delta (fg)=f\delta(g)+g\delta (f)+\delta (f)\delta
(g)$, $\forall f,g\in k[x]$.
\item[(e)] In particular $\delta (fg)=f\delta(g)$ $\forall f\in
\ker (\delta),\forall g\in k[x]$.
\item[(f)] $\delta (p^2)=(\delta (p))^2 $,
$\forall p\in k[x]$. Consequently, $k^+[x]=\ker (\delta)$ is
stable under the endomorphism $p\longmapsto p^2$.
\end{enumerate}\eprop

\brem\label{deg10} 1. By (a) and \ref{plu3}.c below, 
every polynomial $g\in k^+[x]$ 
is of the form $$g=p+p^+=\sum_{0\le k_1<\ldots<k_n}
\left(x^{2k_i+1}+(x+1)^{2k_i+1}\right)=g_1(x^2+x)$$
for some 
odd $p\in k[x]$ and some $g_1\in k[x]$. 

Similarly every polynomial $f$ satisfying $f^+=f+1$ is of the form
$f=x+g$ for some $g\in k^+[x]$. 
For instance, by virtue of \ref{dpo1}.c this is so for the polynomials
$f_q=F_{q-1}+F^+_{q+1}$ $\forall q=2^r$. \erem

We let $\Irr^+[x]=\Irr[x]\cap k^+[x]\subseteq k[x]$ be the set
of all irreducible self-conjugate polynomials.

\bexas\label{deg8} 1. The polynomials in $\Irr^+[x]$ of degree
$\le 8$ are the following ones:
$$x^2+x+1,\quad x^4+x+1,\quad x^6+x^5+x^3+x^2+1,$$
$$x^8+x^6+x^5+x^3+1,\quad x^8+x^6+x^5+x^4+x^3+x+1\,.$$

2. $F_5=(x^2+x+1)^2$ is the only self-conjugate Fibonacci
polynomial \cite{GKW}.  
\eexas

\bsit\label{irr1} For any $q=2^r$, we let $\Tr_q
(x)=\sum_{i=0}^{r-1} x^{2^i}$, so that $\Tr_q (x+y)=\Tr_q
(x)+\Tr_q (y)$. We let $h_r(x)=x^q+x+1\in k[x]$ and $\tilde
h_r(x)=1+\Tr_{q} (x)$. The following proposition can be checked
readily, see \cite[Theorem 3.80]{LN} for (b) and 
\cite[Exercise 3.90]{LN} for (c).\esit

\bprop \label{plu1}
\begin{enumerate}\item[(a)] $\forall q=2^r$, $r\ge 0$, we have
$$h_r(x)=\tilde h_r(x(x+1))=1+\Tr_{\F_q} (x)+\Tr_{\F_q}^2 (x)\in
k^+[x]$$ and $$\tilde h_{2r} (x)=1+\Tr_{q^2} (x) = 1+\Tr_{q}
(x)+\Tr^q_{q} (x)=h_r (\Tr_q (x))\in k^+[x]\,.$$
\item[(b)] 
The  decompositions of $\tilde h_r,\,h_r$ into irreducible factors
over $\F_q$ are, respectively, $$\tilde h_r(x)=\prod_{j=1}^{q/2}
(x+\beta_j)\qquad\mbox{and}\qquad h_r(x)=\prod_{j=1}^{q/2} (x^2+
x+\beta_j)\,,$$ where $\beta_j\in\F_q$ runs over the affine
subspace $\{z\in \F_q\,:\,\Tr_{\F_q} (z)=1\}$.
\item[(c)] $\forall u,v\in \bar k$,
$h_r (u+v)=h_r (u)+h_r (v)+1$. The splitting field of $h_r$ is
$\F_{q^2}$. Moreover $h_r$ has simple roots that fill in the
$r$-dimensional affine subspace
$$\{z\in \F_{q^2}\,:\, z^q=z+1\}=\{z\in \F_{q^2}\,:\, \rho
(z)=\rho_q (z)\}\subseteq \F_{q^2}\,,$$ where $\rho: z\longmapsto z+1$ 
and 
$\rho_q :
z\longmapsto z^q$. This subspace is parallel to
$h_r^{-1} (1)=\F_q$ and stable under the Frobenius automorphism
and under the involutions $\rho$ and $\rho_q$.
\end{enumerate}\eprop

$\forall e\in\N$ and $\forall m\in\NO$, we let $D_2(2^em)=e$.

\bprop \label{plu2}
\begin{enumerate}
\item[(a)] $h_r\circ h_s = h_s\circ h_r = h_{r+s}+h_r+h_s$.
Consequently, $h_{2s}=h_s\circ h_s$. More generally, 
$\forall q=2^r$ and $\forall s\in\N$,
$$h_{qs}=\underset{q}{\underbrace{h_s\circ\ldots\circ h_s}}\,.$$
\item[(b)]$h_s\mid h_{r} \,\,\iff\,\, r=ms$, where
$m\in\NO$. In particular $h_1\mid h_r\,\,\iff\,\, r\in\NO$.
\item[(c)]
$$\gcd (h_r,h_s)=\begin{cases} h_{\gcd (r,s)} &\mbox{if}\quad 
D_2(r)=D_2(s),\\
1 & otherwise.\end{cases}$$ Similarly, $$\gcd (\tilde h_r,\tilde
h_s)=\begin{cases} \tilde
h_{\gcd (r,s)} &\mbox{if}\quad D_2(r)=D_2(s),\\
1 & otherwise.\end{cases}$$
 \end{enumerate}\eprop

 \bproof The proof of (a) is easy and can be omitted.
To show
 (b) we assume that $h_s\mid h_{r}$, and we let $q=2^r,\,q'=2^s$. 
 By virtue of
 \ref{plu1}.b, $\F_{q'^2}\subseteq\F_{q^2}$, hence $s\mid
 r$. If $r=ms$ then for any root $z$ of $h_s$,
 $$z^q=z^{q'^m}=\begin{cases} z & \mbox{if}\,\, m\,\,\,\mbox {is even},\\
 z^{q'}& \mbox{otherwise}.\end{cases}$$ Thus $h_s\mid h_{r}$ implies
 that $m\in\NO$. The converse is easy.

 (c) Assume that $h_s$ and $h_{r}$ have a common root $z\in\bar k$ 
 of degree $\delta$.
 By virtue of \ref{plu1}.c,
 $\F_{2^\delta}\subseteq\F_{2^{2r}}\cap\F_{2^{2s}}$, hence
 $\delta\mid 2\gcd (r,s)$. Since $z^{2^{r}}=z^{2^{s}}=z+1\neq z$,
 we have
 $z\in (\F_{2^{2r}}\setminus \F_{2^{r}})\cap (\F_{2^{2s}}\setminus
 \F_{2^{s}})$. It follows that $\delta$ does not divide $\gcd (r,s)$ i.e.,
 $\delta=2\delta'$, where $\delta'\mid \gcd (r,s)$.

 Letting $r=m\delta'$ and $\tilde q=2^{\delta'}$ we obtain
 $$z+1=z^{2^r}=z^{{\tilde q}^m}=\begin{cases} z & \mbox{if}\,\, 
 m\,\,\,\mbox {is even},\\
 z^{\tilde q}& \mbox{otherwise}.\end{cases}$$ Hence $m\in\NO$.
 Similarly, $s=n\delta'$, where $n\in\NO$, and so,
 $D_2(r)=D_2(s)=D_2(\delta')$. Moreover, $h_{\delta'} (z)=0$. It
 follows that $\gcd (h_r,h_s)\mid h_{\gcd (r,s)}$. Vice versa, 
 by virtue of
 (b),
 $h_{\gcd (r,s)}\mid \gcd (h_r,h_s)$.
  Therefore (c) follows.
 \eproof

\bsit\label{irr2}For a polynomial
$\tau=x^r+a_{r-1}x^{r-1}+\ldots\in \Irr[x]$ we denote $\Tr
(\tau)=a_{r-1}=\Tr_{\F_q} (z)$, where $q=2^r$ and $z\in \F_q$ is
a root of $\tau$.
 We let $$\Irr_i[x]=\{\tau\in \Irr[x]\,:\,
\Tr (\tau)=i\},\quad i=0,1,$$ and $$\Irr^- [x]=\{\tau\tau^+\,:\,
\tau\in \Irr[x]\setminus\Irr^+[x] \}\,.$$ \esit

\bprop \label{plu3}
\begin{enumerate} \item[(a)]
Every irreducible factor of $h_r$ belongs to $\Irr^+[x]$, and
every $\tau\in\Irr^+[x]$ divides one of the $h_r$, $r\ge 1$.
\item[(b)] Every polynomial $\tau\in\Irr^+[x]$ of degree $2r$
admits a decomposition
$$\tau(x)=\prod_{i=0}^{r-1}
(x^2+x+\beta^{2^i})\,,$$ where $\beta\in\F_q$ ($q=2^r$) and $\Tr_{\F_q}
(\beta)=1$.
\item[(c)] The map
$$\alpha: k[y]\to k^+[x]=k[x(x+1)],\qquad q(y)\longmapsto p(x):=q(x(x+1))\,,$$
is an isomorphism of $k$-algebras.
\item[(d)] $\Irr (k^+[x])=\alpha (\Irr[x])=\Irr^+[x]\bigcup \Irr^- [x]\,.$
Moreover $\alpha (\Irr_1[x])=\Irr^+[x]$ and $\alpha (\Irr_0[x])=
\Irr^-[x]$.
\end{enumerate}\eprop

\bproof An element $z\in\bar k$ is a root of one of the
polynomials $h_r$, $r\ge 1$, if and only if $z$ and $z+1$ are
conjugated, if and only if the minimal polynomial $\tau$ of $z$
is stable under the involution $\rho:z\longmapsto z+1$ i.e.,
$\tau\in\Irr^+[x]$. Hence (a) follows.

(b) is immediate from \ref{plu}.b and \ref{plu1}.b.

To show (c) it is enough to establish that $\alpha$ is
surjective. For every $f=\prod_{i=1}^n \tau_i\in k^+[x]$ of
positive degree, the involution $\rho$ acts on the set
$\{\tau_i\}_{i=1,\ldots,n}$ of all irreducible factors of $f$.
Since $f=f^+= \prod_{i=1}^n \tau^+_i$ then either
$\tau_i\in\Irr^+[x]$ or $\tau_i^+=\tau_j\neq \tau_i$ $\forall
i=1,\ldots,n$.

For every $\tilde\tau=\prod_{i=0}^{r-1} (x+z^{2^i})\in \Irr[x]$,
$$(\tilde\tau\tilde\tau^+) (x)=\prod_{i=0}^{r-1} (x+z^{2^i})(x+1+z^{2^i})=
\prod_{i=0}^{r-1} (x^2+x+\beta^{2^i})=\alpha(\tau)(x)\,,$$ where
$\beta=z(z+1)\in \F_q$, $q=2^r$, and $\tau\in k[x]$.

We will show below that for every $\tau\in\Irr^+[x]$,
$\tau=\alpha (\tilde\tau)$ for some $\tilde\tau\in\Irr[x]$. Thus
$f\in k[x(x+1)]$. Hence $k^+[x]=k[x(x+1)]$. Now (c) follows.

(d) Let $\beta\in\F_q$, where $q=2^r$, be a root of a polynomial
$\tau\in \Irr_0[x]$ of degree $r\ge 1$.  Then $\Tr_{\F_q}
(\beta)=0$, hence  $\beta=z(z+1)$ for some $z\in\F_q$
\cite[2.80]{LN}. Therefore
$$\tau (x(x+1))=\prod_{j=0}^{r-1} (x^2+x+z^{2^j}(z+1)^{2^j}))=
\prod_{j=0}^{r-1} (x+z^{2^j})\prod_{j=0}^{r-1} (x+1+z^{2^j})=
(\tilde\tau\tilde\tau^+) (x)\,,$$ where $\tilde\tau
(x)=\prod_{j=0}^{r-1} (x+z^{2^j})\in \Irr[x]$ as $\deg z=r=\deg
\,\tilde\tau$.

Suppose further that $\tilde\tau=\tilde\tau^+$. Then $z^{2^i}=z+1$
for some $i\in\{1,\ldots,r-1\}$. Hence $\beta^{2^i}=\beta$, and
so $\deg\beta\le i<r$, a contradiction. This proves the last
equality in (d) (cf. \cite[Exercise 3.86]{LN}).

For every $\tau\in \Irr^+[x]$ of degree $\deg\tau=2r>0$, the
involution $\rho$ acts on the set
$\{z_j=z^{2^j}\}_{j=0,\ldots,2r-1}$ of roots of $\tau$, where $z$
is one of these roots. If $\rho(z)=z+1=z^{2^i}=z_i$ then $\rho
(z_j)=z_j+1=z_{(i+j)\mod 2r}$, $j=0,\ldots,2r-1$. As $\rho^2=\id$ we
have $j+2i\equiv j\mod 2r\,\,\forall j$, hence $i=r$. Therefore
$$\tau (x)= \prod_{j=0}^{2r-1} (x+z_j)=
\prod_{j=0}^{r-1} (x+z_j)(x+z_j+1)=\tilde\tau (x(x+1))=\alpha
(\tilde\tau) (x) \,,$$ where $$\tilde\tau (x)=\prod_{j=0}^{r-1}
(x+\beta_j)=\prod_{j=0}^{r-1} (x+\beta^{2^j})\in k[x]$$ with
$\beta_j=z_j(z_j+1)$ and $\beta=z(z+1)\in\F_q$, $q=2^r$. Since
$\deg\beta=r=(\deg z)/2$, we have $\tilde\tau (x)\in \Irr[x]$.
Moreover $\Tr_{\F_q} (\beta)=1$ since the polynomial $z^2+z+\beta\in
\F_q[x]$ is irreducible over $\F_q$. Hence
$\tilde\tau\in\Irr_1[x]$. This shows that
$\alpha(\Irr_1[x])=\Irr^+[x]$. Now (d) follows.\eproof

\brems\label{irr3} 1. (\cite[Exercise 3.87]{LN}) For any odd prime $p$
such that $\ord_p \,2=p-1$, $\tau
(x)=\frac{x^p+1}{x+1}\in\Irr_1[x]$ and so,  $\alpha
(\tau)\in\Irr^+[x]$.

2. $h_r\in\Irr^+[x]\quad\iff\quad r\in\{1,2\}$.

3.  $\forall r=2^a$, $h_r$ has $2^{r-a-1}$ irreducible factors of  the
same degree $2r=2^{a+1}$.

4. We have $\tilde h_4 (x)=1+\Tr_{2^4} (x)=\tau\tau^+$, where
$\tau,\,\tau^+\in\Irr [x]$, $\tau=x^4+x^3+1\neq \tau^+$. \erems

\end{document}